\documentclass[prb,aps,twocolumn,superscriptaddress,10pt]{revtex4-2}
\usepackage[utf8]{inputenc}
\usepackage[T1]{fontenc}
\usepackage{color}
\usepackage{amsmath}
\usepackage{amsfonts}
\usepackage{hyperref}
\usepackage{amsthm}
\usepackage{bm}
\usepackage{float}
\usepackage{textcomp}
\usepackage{upgreek}
\usepackage{textgreek}
\usepackage{setspace}
\usepackage[percent]{overpic}
\hypersetup{
  colorlinks = true,
  allcolors= blue,
}
\usepackage{tikz,pgfplots}
\usepackage{blkarray}
\usetikzlibrary{patterns}
\usetikzlibrary{fadings}
\usetikzlibrary{decorations.pathmorphing,patterns}
\tikzset{snake it/.style={decorate, decoration=snake}}
\usetikzlibrary{arrows, decorations.markings}
\pgfplotsset{compat=1.14}
\tikzset{
vecArrow/.style={
  thick,
  decoration={markings,mark=at position
   1 with {\arrow[scale=2,thin]{open triangle 60}}},
  double distance=1.4pt, shorten >= 10.5pt,
  preaction = {decorate},
  postaction = {draw,line width=1.4pt, white,shorten >= 4.5pt}
  },
innerWhite/.style={
  semithick,
  white,
  line width=1.4pt,
  shorten >= 4.5pt
  }
}
\definecolor{orange}{rgb}{1,0.5,0}
\definecolor{darkgreen}{rgb}{0,0.4,0.1}

\newcommand{\WidthFigure}{\columnwidth}

\makeatletter
\newcommand{\doublehat}[1]{%
\begingroup%
  \let\macc@kerna\z@%
  \let\macc@kernb\z@%
  \let\macc@nucleus\@empty%
  \hat{\raisebox{.3ex}{\vphantom{\ensuremath{#1}}}\smash{\hat{#1}}}%
\endgroup%
}
\makeatother

\makeatletter
\newcommand{\doublehatSub}[1]{%
\begingroup%
  \let\macc@kerna\z@%
  \let\macc@kernb\z@%
  \hat{\raisebox{-.07ex}{\vphantom{\ensuremath{#1}}}\smash{\hat{#1}}}%
\endgroup%
}
\makeatother

\makeatletter
\DeclareFontFamily{OMX}{MnSymbolE}{}
\DeclareSymbolFont{MnLargeSymbols}{OMX}{MnSymbolE}{m}{n}
\SetSymbolFont{MnLargeSymbols}{bold}{OMX}{MnSymbolE}{b}{n}
\DeclareFontShape{OMX}{MnSymbolE}{m}{n}{
    <-6>  MnSymbolE5
   <6-7>  MnSymbolE6
   <7-8>  MnSymbolE7
   <8-9>  MnSymbolE8
   <9-10> MnSymbolE9
  <10-12> MnSymbolE10
  <12->   MnSymbolE12
}{}
\DeclareFontShape{OMX}{MnSymbolE}{b}{n}{
    <-6>  MnSymbolE-Bold5
   <6-7>  MnSymbolE-Bold6
   <7-8>  MnSymbolE-Bold7
   <8-9>  MnSymbolE-Bold8
   <9-10> MnSymbolE-Bold9
  <10-12> MnSymbolE-Bold10
  <12->   MnSymbolE-Bold12
}{}

\let\llangle\@undefined
\let\rrangle\@undefined
\DeclareMathDelimiter{\llangle}{\mathopen}%
                     {MnLargeSymbols}{'164}{MnLargeSymbols}{'164}
\DeclareMathDelimiter{\rrangle}{\mathclose}%
                     {MnLargeSymbols}{'171}{MnLargeSymbols}{'171}
\makeatother
\DeclareUnicodeCharacter{300}{à}

\DeclareMathAlphabet{\mathsfit}{\encodingdefault}{\sfdefault}{m}{sl}
\SetMathAlphabet{\mathsfit}{bold}{\encodingdefault}{\sfdefault}{bx}{sl}
\newcommand{\tens}[1]{\bm{\mathsfit{#1}}}
\newcommand{\tenscomp}[1]{\mathsfit{#1}}

\makeatletter
\usepackage{comment}
\let\wfs@comment@comment\comment
\let\comment\@undefined

\usepackage[final]{changes}
\let\wfs@changes@comment\comment
\let\comment\@undefined

\newcommand\comment{%
    \ifthenelse{\equal{\@currenvir}{comment}}
    {\wfs@comment@comment}
    {\wfs@changes@comment}%
}
\makeatother

\begin{document}

\setcitestyle{super}
\title{Thermal conductivity of glasses above the plateau: first-principles theory and applications}

\author{Michele Simoncelli}
\email{ms2855@cam.ac.uk}
\affiliation{ 
Theory of Condensed Matter Group of the Cavendish Laboratory, University of Cambridge (UK)}
\author{Francesco Mauri}
\affiliation{Dipartimento di Fisica, Universit{\`a} di Roma La Sapienza, Italy}
\author{Nicola Marzari}
\affiliation{Theory and Simulation of Materials (THEOS) and National Centre for Computational Design and Discovery of Novel Materials (MARVEL), {\'E}cole Polytechnique F{\'e}d{\'e}rale de Lausanne, Lausanne, Switzerland.}

\begin{abstract}
Predicting the thermal conductivity of glasses from first principles has hitherto been a prohibitively complex problem.
In fact, past works have highlighted challenges in achieving computational convergence 
with respect to length and/or time scales using either
 the established Allen-Feldman or Green-Kubo formulations, endorsing the concept that atomistic models containing thousands of atoms  ---  thus beyond the capabilities of first-principles calculations  ---  are needed to describe the thermal conductivity of glasses.
In addition, these established formulations either neglect anharmonicity (Allen-Feldman) or miss the Bose-Einstein statistics of atomic vibrations (Green-Kubo), thus leaving open the question on the relevance of these effects.
Here, we present a first-principles formulation to address the thermal conductivity of glasses above the plateau, which can account comprehensively for the effects of structural disorder, anharmonicity, and quantum Bose-Einstein statistics. The protocol combines the Wigner formulation of thermal transport with convergence-acceleration techniques, and is validated in vitreous silica using both first-principles calculations and a quantum-accurate  machine-learned interatomic potential.
We show that
models of vitreous silica containing less than 200 atoms can already reproduce 
the thermal conductivity {in the macroscopic} limit and that anharmonicity negligibly affects heat transport in vitreous silica. We discuss the microscopic quantities that determine the trend of the conductivity at high temperature, highlighting the agreement of the calculations with experiments in the temperature range above the plateau where radiative effects remain negligible ($50{\lesssim} T{\lesssim}450$ K).
\end{abstract}

\maketitle

\section{Introduction} 
The thermal conductivity of glasses 
is a key property for many and diverse technological applications, including, e.g., the microelectronics \cite{pasquarello1998interface}, pharmaceutical \cite{niu2018molecular}, aerospace \cite{uyanna2020thermal,Silica_ML}, optic\cite{kotz2017three}, and construction~\cite{arbab2010glass} industries.
To a large degree of universality, for glasses there are three characteristic intervals in the temperature dependence of the thermal conductivity $\kappa(T)$ \cite{Freeman_Anderson,allen1989thermal,allen1993thermal,allen1999diffusons}: 
(i)~the low temperature region ($T{\lesssim}  1\; K$) where $\kappa(T){\sim} T^2$; 
(ii)~the intermediate temperature region ($5 {\lesssim} T  {\lesssim} 25\; K$) where the conductivity displays a plateau ($\kappa(T){\sim} {\rm constant}$);
(iii)~the above-the-plateau region ($T{\gtrsim}30$ K) where the thermal conductivity increases with temperature.
The trend observed in region (i) is usually explained relying on the phenomenological \cite{Freeman_Anderson,leggett2013tunneling,paz2014identification,muller2019towards} ``tunneling two-level system'' (TLS) model \cite{phillips1972tunneling,anderson1972_TLS}. The plateau found in the region (ii) has motivated several studies, which explored its possible connection with the boson peak \cite{Anharmonicity_Boson_peak_glasses,schirmacher2006thermal,lubchenko2003origin,PhysRevB.102.024202}.
The above-the-plateau region (iii)  ---  which is the most relevant one for the aforementioned technological sectors  ---  has up to now been investigated using the Allen-Feldman (AF) formulation~\cite{allen1989thermal} 
or molecular dynamics (MD) methods (the Green-Kubo formulation in combination with classical \cite{McGaughey2009predicting,donadio2009atomistic,Larkin_2014,lv2016_locons}, first-principles \cite{marcolongo2016microscopic,carbogno2017ab,puligheddu2017first,ercole2018ab}, or machine learning \cite{verdi2021thermal} MD simulations (GKMD), non-equilibrium MD (NEMD) \cite{Jund1999,McGaughey2009predicting,tian2017thermal}, or the approach-to-equilibrium MD (AEMD) \cite{Lampin2013,Martin2022}).
Past works relying on these formulations have highlighted two major challenges. First, achieving computational convergence with respect to supercell sizes and simulations times is onerous and problematic \cite{isaeva2019modeling,McGaughey2009predicting,feldman1993thermal}, leading to the conclusion that accurate calculations require models containing thousands of atoms, and simulation times of the order of tens to hundreds of picoseconds; this yields a computational cost that is prohibitively high for direct first-principles calculations.
Second, neither Allen-Feldman nor MD formulations can immediately ensure a comprehensive description of thermal transport, since the former neglects anharmonicity (thus it is in principle accurate only in the low-temperature regime where anharmonicity phases out \cite{allen1993thermal}), and the latter, while  accounting for anharmonicity exactly, misses the quantum Bose-Einstein statistics of atomic vibrations \cite{PhysRevMaterials.3.085401}, relevant at low temperatures.

{The recently developed Wigner formulation \cite{simoncelli2019unified,simoncelli2021Wigner} has shown that the two established microscopic heat-conduction mechanisms for crystals and glasses  ---  \textit{i.e.} the particle-like propagation of phonon wavepackets described by the Peierls-Boltzmann equation \cite{peierls1929kinetischen,peierls1955quantum}, and the
wave-like tunneling mechanisms described by the AF equation\cite{allen1989thermal,allen1993thermal}$^,$
\footnote{here the ``diffuson'' mechanisms described by Allen-Feldman theory is denoted with ``wave-like tunneling'' because it bears some analogies to the electronic Zener interband tunneling \cite{PhysRevB.86.155433}, more details about the terminology can be found in Refs.~\cite{simoncelli2019unified,simoncelli2021Wigner}.}, respectively  ---  both emerge as limiting cases from a unified transport equation. 
Such a Wigner formulation offers a comprehensive approach to describe heat transport in solids, encompassing: (i) ``simple crystals'', characterized by phonon interband spacings much larger than the linewidths, where particle-like propagation dominates and the Peierls-Boltzmann limit is accurate;  (ii) glasses, where wave-like tunneling is relevant and the Wigner formulation extends AF theory accounting for anharmonicity;
(iii) the intermediate regime of ``complex crystals'', where particle-like propagation and wave-like tunneling are simultaneously present and 
neither Peierls-Boltzmann or Allen-Feldman are sufficient. 
This intermediate regime is common and prevalent crystals characterized by phonon interband spacings smaller than the linewidths and featuring ultralow thermal conductivity (e.g. those used in thermoelectrics\cite{PhysRevLett.125.085901} or thermal barrier coatings\cite{simoncelli2021Wigner}). 
It is worth mentioning that the thermal conductivity expression derived from the Wigner formulation has been obtained also following a many-body Green's function approach\cite{isaeva2019modeling,Caldarelli_2022,fiorentino2022green}.}
The Wigner formulation for transport has paved the way, as we will see, to tackle the  aforementioned challenges of comprehensively describing transport in glasses accounting for the interplay between Bose-Einstein statistics, anharmonicity, and disorder. Still, the challenge of predicting the conductivity of glasses from first principles remains, 
due to a structural complexity larger than simple crystals \cite{broido2007intrinsic}, disordered crystals\cite{PhysRevLett.106.045901}, and complex crystals, for which several advances have recently been made  \cite{simoncelli2019unified,xia2020particlelike,PhysRevLett.125.085901,simoncelli2021Wigner,PhysRevB.102.201201}.

Here, we present a protocol that addresses both the aforementioned challenges,
enabling accurate first-principles predictions for the thermal conductivity of glasses, and 
combining the Wigner transport equation\cite{simoncelli2019unified,simoncelli2021Wigner} (WTE) with convergence-acceleration techniques; it is first discussed
in the limiting case of vanishing anharmonicity, where the WTE conductivity reduces to the harmonic AF\cite{allen1989thermal,allen1993thermal} conductivity, and then extended to account for the effects of anharmonicity and evaluate from finite-size models of glasses the bulk limit of the anharmonic WTE conductivity. 
We showcase the protocol in vitreous silica ($v$-SiO$_2$)  ---  a paradigmatic glass that is employed in many and diverse technologies \cite{pasquarello1998interface,kotz2017three,niu2018molecular,Silica_ML,uyanna2020thermal,arbab2010glass}  ---  extending the simulations to very large cell sizes thanks to the use of machine learned potentials (the recently developed Gaussian Approximation Potential\cite{PhysRevLett.104.136403} (GAP) for silica polymorphs \cite{erhard2022machine}).
The GAP potential we employ has been generated from first-principles calculations performed using the SCAN functional\cite{PhysRevLett.115.036402}, and yields quantum-accurate predictions for the vibrational properties of various silica polymorphs\cite{erhard2022machine}.  
Here, we employ GAP to describe the vibrational properties and evaluate the thermal conductivity  ---  both in the harmonic Allen-Feldman or in the anharmonic Wigner framework  ---  of models of $v$-SiO$_2$ having very different sizes, showing how our protocol allows to accurately evaluate  the bulk limit of the harmonic or anharmonic conductivity using models containing less than 200 atoms, 
that are in very good agreement with the {macroscopic} cell limit of 5184 atom \cite{erhard2022machine}.

After having validated the protocol, we employ it to study the conductivity of $v$-SiO$_2$ fully from first principles using models of $v$-SiO$_2$ containing 108\cite{charpentier2009first}, 144\cite{PhysRevB.79.064202,Materials_cloud_structure}, and 192 atoms\cite{Kroll_2013}.
We discuss how the AF or WTE conductivities change if the widely used semi-empirical van Beest-Kramer-van Santen\cite{PhysRevLett.64.1955,carre2007amorphous} (BKS) potential, or the state-of-the-art GAP potential for silica polymorphs \cite{erhard2022machine}, are used in place of first-principles calculations to compute the vibrational properties 
of $v$-SiO$_2$. 
We elucidate analogies and differences between the harmonic  and the anharmonic conductivity, relying on the latter to show how the high-temperature trend of  $\kappa(T)$  
is determined by the variation of the velocity-operator elements, whose value regulates the amount of heat transferred by wave-like tunneling between vibrational eigenstates,
when the energy difference between coupled eigenstates increases. 
Finally, we rationalize these results at the microscopic level, relying on Allen-Feldman's harmonic diffusivity and extending such quantity to account for anharmonicity, but also showing that anharmonicity has negligible effects on the conductivity and diffusivity of $v$-SiO$_2$.

\section{Wigner formulation of thermal transport} 
\label{sec:Wigner_formulation}
As anticipated, the WTE formalism is general and can be used to describe solids with variable degrees of disorder, ranging from ordered crystals with small primitive cells, to disordered glasses with diverging primitive cell (in this latter case, for sufficiently large primitive cells periodic boundary conditions become irrelevant and the Brillouin zone (BZ) reduces to the point $\bm{q}{=}\bm{0}$ only). 
In practice, in numerical simulations non-periodic glasses can be approximatively described as crystals with large but finite primitive cells, thus having a small but finite BZ that includes wavevectors different from $\bm{q}{=}\bm{0}$. We will discuss in the next section the conditions under which non-periodic glasses can be simulated in periodic-boundary conditions,
after having introduced here the salient features of the WTE formulation in the general case where vibrations can have a wavevector different from $\bm{q}{=}\bm{0}$.      

For systems with low conductivity, which are the focus of this work, the WTE can be solved accurately within the so-called single-mode relaxation-time approximation (SMA, see Methods for details), which allows to write the conductivity in the following compact form:
\begin{equation}
\begin{split}
\kappa{=}\frac{1}{\mathcal{V}{N_{\rm c}} }{\sum_{\bm{q},s,s'}}&\!
\frac{\omega(\bm{q})_s{+}\omega(\bm{q})_{s'}}{4}\!\left(\frac{C(\bm{q})_{s}}{\omega(\bm{q})_{s}}{+}\frac{C(\bm{q})_{s'\!}}{\omega(\bm{q})_{s'\!}}\right)\!
\frac{\rVert\tens{v}(\bm{q})_{s,s'}\lVert^2}{3}\\
\times&\pi\mathcal{F}_{[\Gamma(\bm{q})_s{+}\Gamma(\bm{q})_{s'}]}(\omega(\bm{q})_s-\omega(\bm{q})_{s'})\;,
\label{eq:thermal_conductivity_combined}
\end{split}
\raisetag{5mm}
\end{equation}
where the wavevector $\bm{q}$  and the mode index $s$ label a vibrational eigenstate having energy
$\hbar\omega(\bm{q})_s$, anharmonic linewidth $\hbar\Gamma(\bm{q})_s$, and specific heat
\begin{equation}
C(\bm{q})_s{=}C[\omega(\bm{q})_s]{=}\frac{\hbar^2\omega^2(\bm{q})_s }{k_{\rm B} T^2} {\tenscomp{N}}(\bm{q})_s\big({\tenscomp{N}}(\bm{q})_s{+}1\big)  
\label{eq:quantum_specific_heat_A}
\end{equation}
(${\tenscomp{N}}(\bm{q})_s{=}[\exp(\hbar \omega(\bm{q})_s/k_{\rm B}T){-}1]^{-1}$ is the Bose-Einstein distribution at temperature $T$); 
the quantity
\begin{equation}
\vspace*{-0.5mm}
  \rVert\tens{v}(\bm{q})_{s,s'}\lVert^2{=}\sum_{\alpha=1}^3\tenscomp{v}^{\alpha}(\bm{q})_{s,s'}\tenscomp{v}^{\alpha}(\bm{q})_{s',s}\vspace*{-0.5mm}
\end{equation} 
denotes the square modulus of the velocity operator \cite{simoncelli2021Wigner} between eigenstates $s$ and $s'$ at the same wavevector $\bm{q}$ ($\alpha$ in these expressions denotes a Cartesian direction, and since vitreous solids are in general isotropic, the scalar conductivity~(\ref{eq:thermal_conductivity_combined}) is computed as the average trace of the tensor $\kappa^{\alpha\beta}$ given by Eq.~(47) of Ref.~\cite{simoncelli2021Wigner}; $N_{\rm c}$ is the number of $\bm{q}$-points entering in such a summation and $\mathcal{V}$ is the primitive-cell volume).
Finally, $\mathcal{F}$ is a Lorentzian distribution having a full width at half maximum (FWHM) equal to $\Gamma(\bm{q})_s{+}\Gamma(\bm{q})_{s'}$: 
\begin{equation}
\begin{split}
    &\mathcal{F}_{[\Gamma(\bm{q})_s{+}\Gamma(\bm{q})_{s'}]}\big(\omega(\bm{q})_s{-}\omega(\bm{q})_{s'}\big)\\
    &=
  \frac{1}{\pi}\frac{\frac{1}{2}\big(\Gamma(\bm{q})_s+\Gamma(\bm{q})_{s'}\big)}{\big(\omega(\bm{q})_s-\omega(\bm{q})_{s'}\big)^2+\frac{1}{4}\big(\Gamma(\bm{q})_s+\Gamma(\bm{q})_{s'}\big)^2}\;.
\end{split}
  \label{eq:Lorentzian}
\end{equation}
In a crystal the primitive cell and the BZ have a finite volume and can be univocally chosen relying on crystallographic conditions \cite{hinuma2017band}; wavevectors are good quantum numbers and phonon group velocities are well defined. Under these circumstances, it is useful to rewrite the WTE conductivity~(\ref{eq:thermal_conductivity_combined}) as sum of two terms, $\kappa{=}\kappa_P{+}\kappa_C$;
specifically, the term $\kappa_P$ (referred to as ``populations conductivity''\cite{simoncelli2021Wigner,simoncelli2019unified}) is determined by the diagonal ($s{=}s'$) or perfectly degenerate ($s{\neq}s'$ with $\omega(\bm{q})_{s}{=}\omega(\bm{q})_{s'}$) terms  in the summation in the conductivity~(\ref{eq:thermal_conductivity_combined});  in crystals such a term describes the 
Peierls-Boltzmann particle-like heat conduction {(averaged over the spatial directions), since the average trace of the  Peierls-Boltzmann conductivity tensor} can be written as $\kappa_P{=}\tfrac{1}{3}\sum_\alpha \kappa^{\alpha\alpha}_P$ with $\kappa_P^{\alpha\alpha}{=}\frac{1}{\mathcal{V} N_C }\sum_{\bm{q}s}C[\omega(\bm{q})_{s}]{\tenscomp{v}^{\alpha}(\bm{q})_{s,s}}\Lambda^\alpha(\bm{q})_s$, \textit{i.e.} as particle-like vibrations having absolute energy $\hbar\omega(\bm{q})_{s}$ (thus specific heat $C[\omega(\bm{q})_{s}]$) and propagating between collisions over a length\footnote{Here we have exploited the possibility to diagonalize at least one Cartesian components ${\alpha}$ of the velocity operator in the degenerate subspace \cite{fugallo2013ab,simoncelli2021Wigner}, thus ${\tenscomp{v}^{\alpha}(\bm{q})_{s,s}}$ is the propagation velocity in direction $\alpha$ and ${\tenscomp{v}^{\alpha}(\bm{q})_{s,s'}}{=}0$ for $s{\neq} s'$ in the degenerate subspace} $\Lambda^\alpha(\bm{q})_{s}{=}{\tenscomp{v}^{\alpha}(\bm{q})_{s,s}}[\Gamma(\bm{q})_s]^{-1}$.
Conversely, non-degenerate off-diagonal elements (``coherences''\cite{simoncelli2021Wigner,simoncelli2019unified}) account for a different ``Wigner'' conduction mechanisms: they do not have an absolute energy akin to that of a particle-like excitation, but are characterized by an energy difference $\hbar\omega(\bm{q})_{s}{-}\hbar\omega(\bm{q})_{s'}$ and describe a wave-like tunneling conduction mechanisms akin to the electronic Zener interband tunnelling \cite{PhysRevB.86.155433}.
It has been shown in Refs. \cite{simoncelli2019unified,simoncelli2021Wigner,PhysRevX.10.011019} that in simple crystals 
particle-like mechanisms dominate and thus $\kappa_{_{\rm P}}{\gg}\kappa_{_{\rm C}}$, while in complex crystals 
both these mechanisms are relevant, and  $\kappa_{_{\rm P}}$ and $\kappa_{_{\rm C}}$ are of the same order. 
Finally, Refs. \cite{simoncelli2019unified,simoncelli2021Wigner} have shown that in the ordered limit describing a harmonic glass\cite{allen1989thermal,allen1993thermal}, Eq.~(\ref{eq:thermal_conductivity_combined}) reduces to the AF formula for the conductivity of glasses.
Specifically, such an ordered limit requires first to describe a (structurally stable\footnote{With ``structurally stable glass'' we mean a glass in which atoms vibrate around equilibrium positions, thus atomic diffusion is negligible. See e.g. refs.~\cite{Egami2019,moon2021examining,Ruta2014,Ross2014,Buchenau1986,Song2019,yu2013beta,PhysRevB.105.014110} for a discussion of how network structure and chemical composition affect the structural stability of glasses and related phenomena.}) glass as the limiting case of a disordered but periodic crystal with an increasingly larger primitive cell (\textit{i.e.} $\mathcal{V}{\to}\infty$ and thus with the BZ reducing to the point $\bm{q}{=}\bm{0}$ only \cite{allen1989thermal,allen1993thermal}), and then letting each linewidth go to the same infinitesimal broadening\cite{allen1989thermal,allen1993thermal} $\hbar\eta$, $\hbar\Gamma(\bm{q})_s{\to}\hbar\eta{\to} 0$, $\forall \;s$ and $\bm{q}{=}\bm{0}$. 
Under these ideal circumstances  
only $\bm{q}{=}\bm{0}$ is considered in the sum in Eq.~(\ref{eq:thermal_conductivity_combined}), and the
Lorentzian distribution~(\ref{eq:Lorentzian}) becomes a Dirac $\delta$,
\begin{equation}
\begin{split}
    \lim_{\eta{\to}0}\!\!\left[\lim_{\mathcal{V}\to \infty}\mathcal{F}_{[2\eta]}(\omega(\bm{q})_s{-}\omega(\bm{q})_{s'})\right]
   {=}\delta\left({\omega(\bm{q})_s{-}\omega(\bm{q})_{s'} }\right),
\end{split}
   \label{eq:Gaussian}
\end{equation}
implying that the WTE conductivity~(\ref{eq:thermal_conductivity_combined}) reduces exactly to the AF formula for the conductivity of glasses (Eq.~(3) of Ref.\cite{allen1989thermal}).
In practice, this ideal bulk-glass limit cannot be reached in numerical calculations, and anharmonic linewidths strongly vary with temperature. In the next sections we first discuss a protocol that allows to accurately describe glasses at the AF harmonic level using finite-size models, and then we rely on the WTE to extend such a protocol accounting for anharmonicity.

\section{Describing glasses with finite-size models} 
\label{sec:glasses_in_periodic_boundary_conditions}
\subsection{Protocol to evaluate the harmonic Allen-Feldman conductivity} 
\label{sub:simulating_glasses_with_allen_feldman_theory_}
From a microscopic viewpoint, AF conductivity  (Eq.~(\ref{eq:thermal_conductivity_combined}) and (\ref{eq:Gaussian}) with $\bm{q}=\bm{0}$ only) 
describes heat transport in an ideal bulk glass as being mediated by a transfer of energy between vibrations that are degenerate in energy (see Eq.~(\ref{eq:Gaussian})) and not localized in the Anderson sense\cite{Anderson_localization} (this requirement is needed to have non-zero velocity operator elements in Eq.~(\ref{eq:thermal_conductivity_combined})).
In actual calculations a glass is approximately described as a crystal having a primitive cell containing a large but finite number of atoms $N_{\rm at}$.
Such an approximation has two important implications.
First, the BZ corresponding to the (large) finite-size model does not reduce to $\bm{q}{=}\bm{0}$ only but has a (small) finite volume.
Second, in a realistic finite-size model of a glass perfectly degenerate vibrational modes are supposed to be absent (this because exact degeneracies are due to point-group symmetries\cite{RevModPhys_Maradudin68}, which are not present in amorphous systems);
thus, the vibrational spectrum is characterized by an average spacing between vibrational energy levels equal to
\begin{equation}
  \hbar\Delta\omega_{\rm avg}=\frac{  \hbar\omega_{\rm max}}{{3N_{\rm at}}},
  \label{eq:average_spacing}
\end{equation}
where $\hbar\omega_{\rm max}$ is the maximum vibrational energy of the solid (which strongly depends on the chemical composition and negligibly depends on disorder, see Fig.~\ref{fig:lw_compare} in Methods) and $3N_{\rm at}$ is the number of vibrational modes at a fixed $\bm{q}$ point.
Eq.~(\ref{eq:average_spacing}) underlines how degenerate vibrational modes determining the harmonic conductivity emerge as a consequence of disorder, since increasing the accuracy in the description of disorder (\textit{i.e.} increasing $N_{\rm at}$) yields a decrease of the average energy-level spacing, implying that in the ideal limit of a bulk glass ($N_{\rm at}{\to}\infty$) adjacent vibrational eigenstates become degenerate  (hereafter we will use the term ``quasi-degenerate'' to refer to these adjacent vibrational eigenstates that  become degenerate in the ideal limit of a bulk glass).

These two properties offer important insights on how to evaluate the strength of finite-size effects in glasses and consequently extrapolate from the finite-size ``reference cell'' of the model the behavior of the ideal (infinite) glass.
Specifically, the presence of a BZ with a small but finite volume implies that Fourier interpolations can be used to sample vibrations in a $n{\times}n{\times}n$ supercell of the finite periodic model. This procedure simulates a system where vibrations are commensurate to a system that is $n{\times}n{\times}n$ times larger than the reference cell used, but where the disorder length scale remains limited by the size of the reference cell. 
Within this scheme one can obtain information about finite-size effects in multiple ways: (i) for a given finite model, one can study the differences between a calculation performed at $\bm{q}{=}\bm{0}$ only, and a calculation on a $n{\times}n{\times}n$  $\bm{q}$-mesh; (ii) one can repeat the analysis at the previous point employing models having larger and larger reference cells  ---  for sufficiently large models one expects to achieve computational convergence, with the calculation at $\bm{q}{=}\bm{0}$ only and that on the $\bm{q}$-mesh yielding indistinguishable results. 
Moreover, in order to retain the key physical property that couplings between quasi-degenerate eigenstates can occur and contribute to heat transport, the Dirac $\delta$~(\ref{eq:Gaussian}) needs to be replaced with a smooth distribution having a broadening $\eta$ of the order of the average energy-level spacing\cite{allen1989thermal,allen1993thermal,feldman1993thermal} $\hbar\Delta\omega_{\rm avg}$; otherwise, no couplings would take place in any system represented by a finite-size periodic supercell.
 Within such a numerical scheme $\eta$ is just a computational parameter, and as such one expects to find a range of values for $\eta$ (the domain of convergence) for which the conductivity is independent from $\eta$.
The considerations above show that the calculation of the harmonic AF conductivity in finite-size models bears some procedural analogies to electronic-structure calculations in metals,
where the BZ is integrated using a discrete mesh and the Dirac $\delta$ identifying the Fermi level is broadened with an aptly chosen smearing function \cite{PhysRevB.28.5480,PhysRevB.40.3616,PhysRevLett.82.3296}, and one needs to identify a range of ``converged values'' for the smearing and for the size of the BZ-integration mesh for which quantities such as the total energy are practically independent from the smearing. 

\begin{figure}[t]
  \centering
  \includegraphics[width=\WidthFigure]{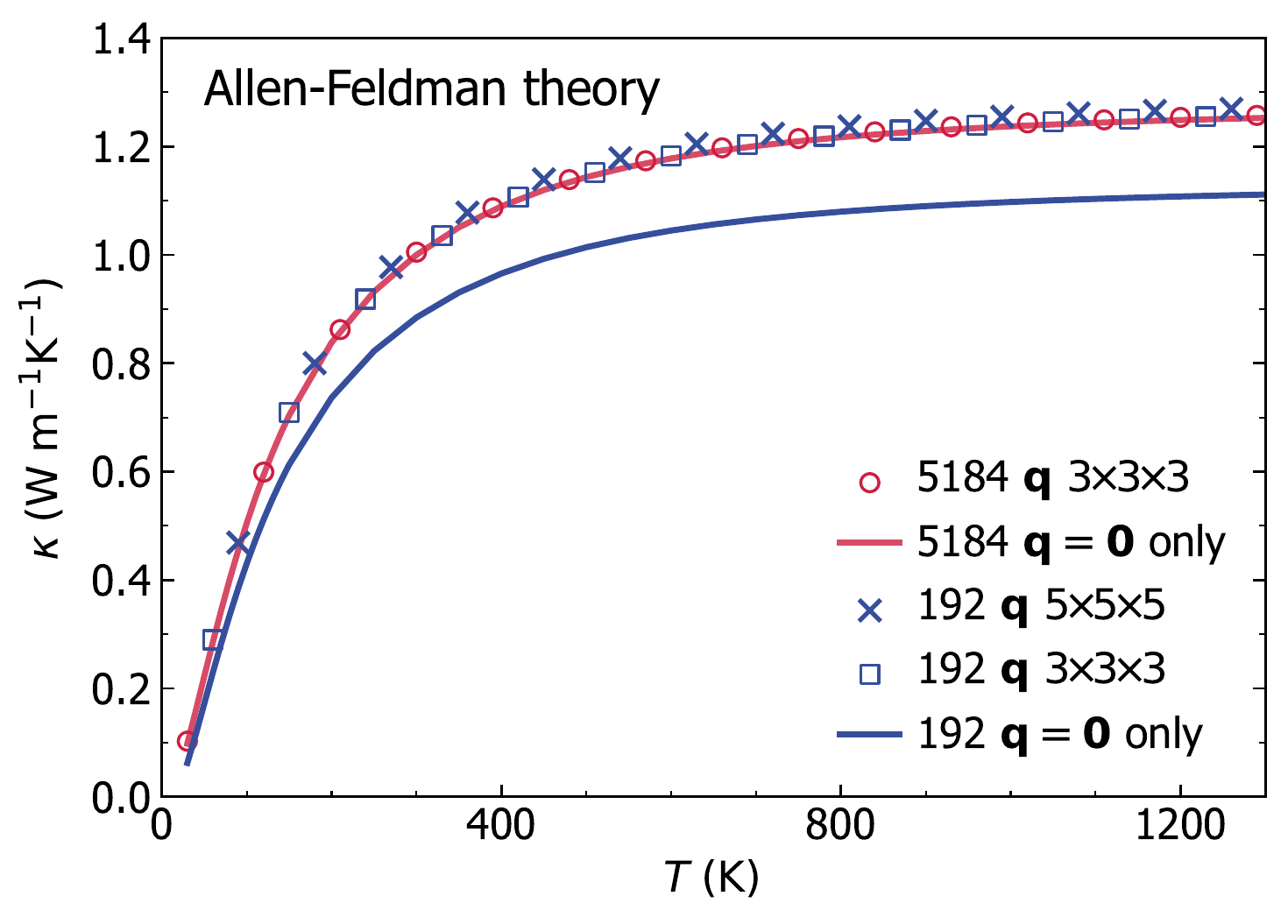}\\[-5mm]
  \caption{
\textbf{Convergence of the Allen-Feldman theory in finite-size  $v$-SiO$_2$ models.} Solid lines are the AF conductivities computed at $\bm{q}{=}\bm{0}$ only, empty circles and squares (crosses) are the AF conductivities computed using the Fourier interpolation on a $3{\times}3{\times}3$ ($5{\times}5{\times}5$) $\bm{q}$ mesh; results for a 192-atom (5184-atom) model are in blue (red). 
All calculations are performed using the same Gaussian broadening $\eta{=}4\;\rm{cm}^{-1}$ for the Dirac $\delta$ appearing in the AF conductivity expression (such a value corresponds to a FWHM larger than the average energy-level spacing of these models, and for this broadening computational convergence is achieved in both structures, see text and Fig.~\ref{fig:harm_theory_plateau}). 
The discrepancy between the solid blue line and all the other data shows that the Fourier interpolation is necessary to achieve computational convergence using a small 192-atom model; the good agreement between scatter points and the red line shows that computational convergence can be achieved with a calculation at $\bm{q}{=}\bm{0}$ only for the larger 5184-atom model.  }
  \label{fig:silica_AF_GAP}
\end{figure}

Pioneering work\cite{allen1989thermal,allen1993thermal,allen1999diffusons,feldman1993thermal} evaluated the AF conductivity broadening the Dirac $\delta$~(\ref{eq:Gaussian}) with a fat-tailed\footnote{We employ the nomenclature used in statistics, which classifies as ``light-tailed'' (``fat-tailed'') all the distributions having tails smaller (larger) than the exponential distribution.} Lorentzian distribution having FWHM of the order of $\hbar\Delta\omega_{\rm avg}$, limited calculations to $\bm{q}{=}\bm{0}$  to reduce the computational cost, and relied on smoothness arguments \cite{allen1993thermal} to extrapolate the bulk limit from finite-size models.
Nevertheless, recent work \cite{McGaughey2009predicting,isaeva2019modeling,PhysRevB.106.014305} have highlighted challenges in achieving computational convergence following such computational protocol. 
Therefore, here we take inspiration from the computational techniques alluded to above for electronic-structure calculations, to tackle the problem of achieving convergence in the calculation of the AF conductivity, using vitreous silica as a paradigmatic test case.
We first show in Fig.~\ref{fig:silica_AF_GAP} that state-of-the-art models of vitreous silica\cite{erhard2022machine} ($v$-SiO$_2$) containing 192 or 5184 atoms
(both studied using a recently developed GAP potential\cite{erhard2022machine}, see Methods for details) 
yield bulk AF conductivities that are perfectly compatible, provided these are computed using a light-tailed Gaussian broadening having a FWHM larger than the average energy-level spacing (more details on the convergence domain for the broadening parameter are discussed later), and a Fourier interpolation is used to extrapolate the bulk limit from the small 192-atom model (the Fourier interpolation has a negligible effect on the large 5184-atom model, confirming the aforementioned expectations).
\begin{figure*}
  \centering
\includegraphics[width=\textwidth]{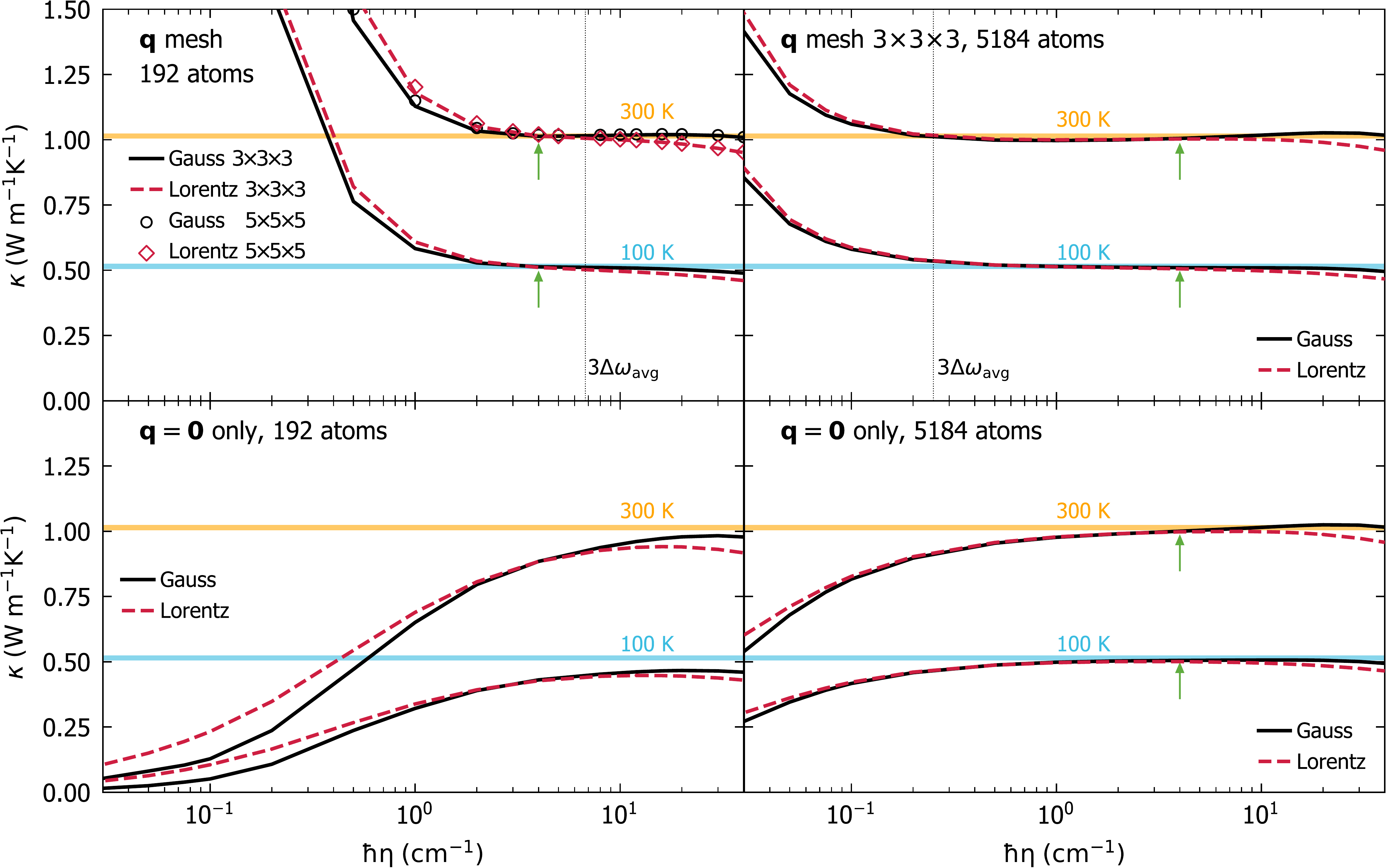}\\[-3mm]
  \caption{\textbf{Convergence of the AF conductivity for $v-$SiO$_2$ with respect to the broadening $\eta$ for the Dirac $\delta$,} for a 192-atom model (left) and for a 5184-atom model (right). Dashed red lines represent a fat-tailed Lorentzian having FWHM $2\eta$, solid black lines represent a light-tailed Gaussian with variance $\eta^2{\pi/2}$ (\textit{i.e.} the same maximum $(\pi\eta)^{-1}$ of the Lorentzian with FWHM $2\eta$). 
  The top row shows that evaluating the AF conductivity using the $\bm{q}$-interpolation technique (see text) and the Gaussian broadening yields a range of value for $\eta$ for which the AF conductivity is not sensitive to the value of $\eta$ (\textit{i.e.} a ``convergence plateau'') for both the 192- and 584-atom models. The horizontal orange and blue lines show the computationally converged value for the AF conductivity at 100 and 300 K, respectively; we highlight that the 192- and 5184-atom models yield compatible (indistinguishable) conductivities.
  In general, but especially in the 192-atom model, the Gaussian broadening yields a wider and more clear convergence plateau compared to the Lorentzian broadening; thus, from a numerical viewpoint, the former has some advantage over the latter. The vertical lines show that when the $\bm{q}$-interpolation technique is adopted, a broadening $\eta=3\Delta\omega_{\rm avg}$ yields a computationally converged thermal conductivity at both 100 and 300 K. 
In the bottom line we show that the evaluation of the AF conductivity at $\bm{q}{=}\bm{0}$ only yields in the large 5184-atom model a convergence plateau that is smaller compared to that obtained in the same structure using $\bm{q}$-interpolation;  at $\bm{q}{=}\bm{0}$  the small 192-atom model instead does not show clear convergence with respect to the broadening. 
The green arrows show the value of $\eta=4$ cm$^{-1}$, which yields a computationally converged conductivity  
in the 192-atom model in the $\bm{q}$-mesh calculation, as well as in 5184-atom model in both the $\bm{q}$-mesh and $\bm{q}{=}\bm{0}$ only calculations.
{The origin of the opposite trend of the broadening-conductivity curve obtained using $\bm{q}$ interpolation or $\bm{q}{=}\bm{0}$ only is discussed in Sec.~\ref{sub:Protocol_to_evaluate_the_anharmonic_Wigner_conductivity}.}
}
  \label{fig:harm_theory_plateau}
\end{figure*}
The details on the domain of convergence for the broadening parameter $\eta$ and 
mesh used for the BZ sampling are reported in Fig.~\ref{fig:harm_theory_plateau}, where we show that computational convergence can be achieved for both the 192- and 5184-atom models, \textit{i.e.} there exists a range of values for $\eta$ over which the conductivity is not sensitive to the value of $\eta$ (``convergence plateau'').
In the larger 5184-atom model the convergence plateau extends to smaller values of $\eta$; this in agreement with the expectation, based on Eq.~(\ref{eq:average_spacing}), that larger models allow for more accurate approximations of the Dirac $\delta$. Importantly, we also show that the Gaussian broadening yields a wider convergence plateau compared to the Lorentzian broadening, especially at low temperatures where anharmonicity phases out and thus the harmonic AF theory is accurate. 
{We note in passing that the improved computational performance found here for the 
Gaussian broadening compared to the 
Lorentzian broadening bears analogies to density-functional-theory calculations for metals, 
where refined representations of the Dirac delta \cite{Marzari_1996,PhysRevLett.82.3296,PhysRevB.28.5480,PhysRevB.40.3616,PhysRevLett.82.3296} are used in place of the Fermi-Dirac thermal broadening \cite{Degio_smearing}
to improve convergence with respect to Brillouin-zone sampling.} 
The analysis of Fig.~\ref{fig:harm_theory_plateau} demonstrates that the evaluation of the AF conductivity at $\bm{q}{=}\bm{0}$ only does not show clear convergence with respect to the broadening for the 192-atom model; a convergence plateau emerges instead if the Fourier interpolation is employed to extrapolate to the bulk limit. 
In contrast, in the 5184-atom model computing the AF conductivity at $\bm{q}=\bm{0}$ only or using an interpolation mesh does not produce significant differences, provided a value of $\eta$ belonging to the convergence plateau is used.
This shows that the protocol of using the
$\bm{q}$-mesh interpolation and (ideally) Gaussian broadening to compute the AF conductivity allows to achieve computational convergence in three small models (containing $\lesssim 200$ atoms) typically affordable in first-principles studies; in $v$-SiO$_2$ the model containing 5184 atoms already at $\bm{q}=\bm{0}$ is accurate,  and allows to determine a convergence plateau for the broadening $\eta$. 

\subsection{Extension of the protocol to evaluate the anharmonic Wigner conductivity} 
\label{sub:Protocol_to_evaluate_the_anharmonic_Wigner_conductivity}
\begin{figure*}
  \centering
  \includegraphics[width=\textwidth]{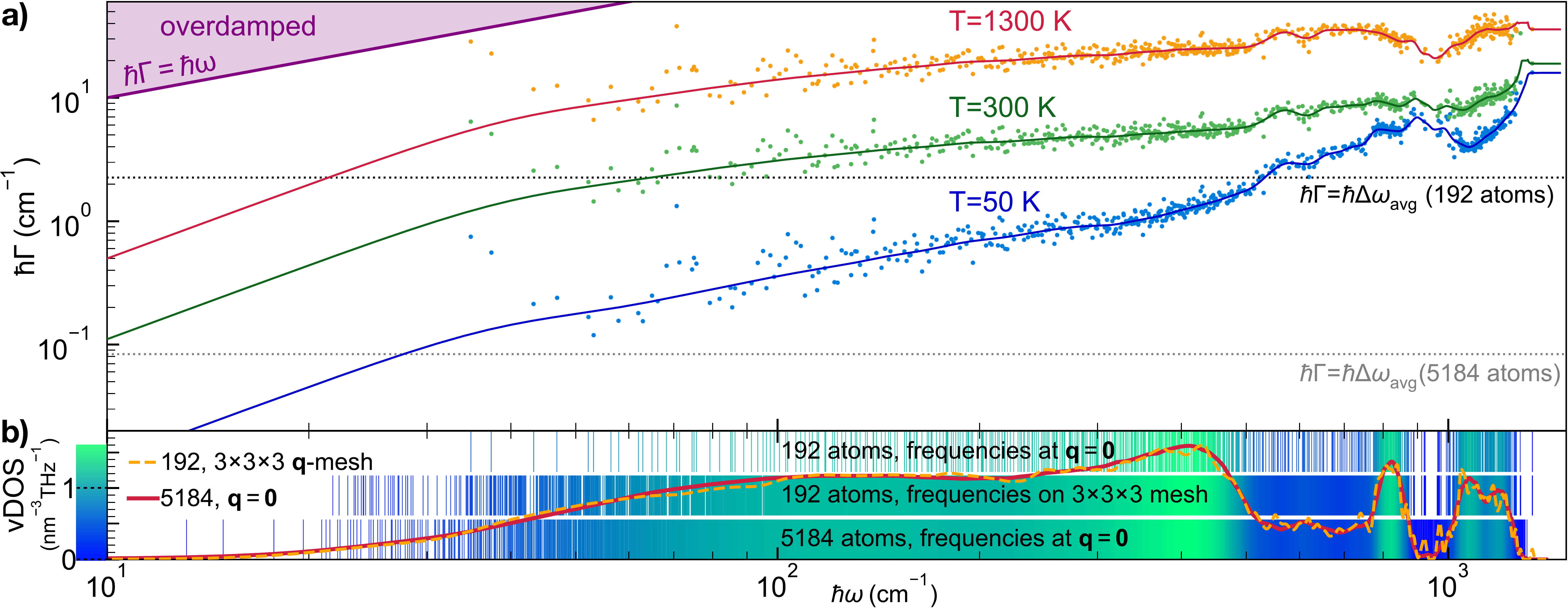}\\[-3mm]
  \caption{
\textbf{Vibrational frequencies and anharmonic linewidths of vitreous silica.} 
Panel \textbf{a)}: the scatter points represents the linewidths as a function of frequency and temperature for the 192-atom model of $v-$SiO$_2$; these are computed at $\bm{q}{=}\bm{0}$ only accounting for third-order anharmonicity\cite{paulatto2015first,paulatto2013anharmonic,li2014shengbte,carrete2017almabte,phono3py} and scattering due to isotopic mass disorder at natural abundance\cite{tamura1983isotope}. 
The solid lines represent a coarse graining of the frequency-linewidth distribution into a single-valued function\cite{garg2011thermal_PhD,PhysRevB.105.134202} $\Gamma_a[\omega]$   (see Methods for details).
The horizontal black (gray) lines show the average energy-level spacing for the 192-atom (5184-atom) model; for the 192-atom (5184-atom) model all the vibrations below the black (gray) line are not accurately accounted for by the bare WTE~(Eq.~(\ref{eq:thermal_conductivity_combined}) and (\ref{eq:Lorentzian})), and need to be regularized using the protocol described in the main text.
Panel \textbf{b)}: the solid red line represents the vibrational density of state (vDOS) computed using the 5184-atom model and considering $\bm{q}{=}\bm{0}$ only, the overlapping dashed orange line is the vDOS computed using the 192-atom model over a $3{\times}3{\times}3$ $\bm{q}$-mesh (both these calculations use the same Gaussian broadening $\eta=4\rm{cm}^{-1}$, discussed also in Figs.~\ref{fig:silica_AF_GAP},\ref{fig:harm_theory_plateau}).
The vertical lines show how the vibrations of $v-$SiO$_2$ are sampled using the 192-atom model at $\bm{q}{=}\bm{0}$ only (top, corresponding to the abscissas of the scatter points in panel \textbf{a)}), using the 192-atom model and relying on Fourier interpolation on a $3{\times}3{\times}3$ $\bm{q}$-mesh (center), and using the 5184-atom model at $\bm{q}{=}\bm{0}$ only (bottom).
  }
  \label{fig:linewidths_sampling}
\end{figure*}

We start by recalling that the WTE also generalizes the AF model accounting for anharmonicity; in particular, the Lorentzian distribution~(\ref{eq:Lorentzian}) appearing in the WTE conductivity~(\ref{eq:thermal_conductivity_combined}) has a FWHM determined by the anharmonic linewidths.
From a computational perspective, and recalling that the size of the model determines the average energy-level spacing~(\ref{eq:average_spacing}),
one expects that evaluating the WTE using a finite-size model should lead to results negligibly affected by finite-size effects whenever all the anharmonic linewidths of the finite-size model are larger than its average energy-level spacing. 
To proceed, we first show in Fig.~\ref{fig:linewidths_sampling}\textbf{a)} the frequency-linewidth distributions for the 192- and 5184-atom $v-$SiO$_2$ models, and compare these with the average energy-level spacing. Each cloud of points represents the frequency-linewidth distribution at a given temperature for the 192-atom model, evaluated explicitly at $\bm{q}=\bm{0}$ only and accounting both for third-order anharmonicity\cite{paulatto2015first,paulatto2013anharmonic,li2014shengbte,carrete2017almabte,phono3py} and for scattering due to isotopic mass disorder at natural abundance\cite{tamura1983isotope}  (see Methods for details). Each solid line is a 
coarse-grained interpolation of the frequency-linewidth distribution into a single-valued function $\Gamma_a[\omega]$; such a coarse graining is inspired by past work \cite{garg2011thermal_PhD,PhysRevB.105.134202} and validated in  Fig.~\ref{fig:interpolation} in the Methods, where we show that evaluating the WTE conductivity using the exact frequency-linewidth distribution or linewidths determined using the coarse grained function $\Gamma(\bm{q})_s{=}\Gamma_a[\omega(\bm{q})_s]$ yields practically indistinguishable results. 
Clearly, in the 192-atom model a significant portion of the vibrational modes have linewidths below the average energy-level spacing at room temperature and below (this happens both when vibrations are sampled at $\bm{q}{=}\bm{0}$ only, as well as when they are more densely sampled on a $3\!{\times}\!3\!{\times}\!3$ $\bm{q}$-mesh).
For the 5184-atom model, instead, the average energy-level spacing is more than one order of magnitude smaller than that of the 192-atom model, implying that already at 50 K most of the linewidths are larger than the average energy-level spacing.

To better understand the combined effect of anharmonicity and of the finite size of the model on conductivity, we recall 
once again that when the WTE conductivity (\ref{eq:thermal_conductivity_combined}) and (\ref{eq:Lorentzian}) is evaluated repeating periodically a finite-size {reference cell}, the conductivity can be decomposed in the sum of a populations and a coherences contribution ($\kappa_{\rm P}$ and $\kappa_{\rm C}$, respectively, see Sec.~\ref{sec:Wigner_formulation}).
{We stress that the populations (coherences) conductivity can be rationalized in terms of a particle-like (wave-like) conduction mechanisms exclusively in actual crystals, where the periodicity and symmetries of the atomic structure imply that the BZ and the vibrational spectrum can be defined unambiguously\cite{hinuma2017band};
consequently, the group velocities with which the particle-like phonon wavepackets propagate, and the interband spacings that characterize the wave-like tunneling of phonons, are well defined.
In finite-size glass models, instead, the decomposition in populations and coherences conductivities is useful to understand finite-size effect 
(more on this later), but cannot be further interpreted in terms of particle-like and wave-like conduction mechanisms.
In fact, such an interpretation is well defined only when the BZ and vibrational spectrum can be unambiguously defined, while in finite-size models of glasses the BZ volume decreases as the size of the model increases.
Finally, we also note that the distinction between populations and coherences conductivities in finite-size models of glasses is different from the distinction between ``propagons'' and ``diffusons'' conductivities discussed by Allen \textit{et al.}\cite{allen1999diffusons} for an ideal (infinitely large) glass. The propagons and diffusons conductivities differentiate the contributions to heat transport originating from vibrational modes that propagate ballistically (akin to phonon wavepackets) from those that do not in an ideal glass; future work will aim at decomposing the total Wigner conductivity into those contributions. }

In finite-size models of glasses the populations conductivity $\kappa_{\rm P}$ is given by the terms diagonal in the mode index ($s{=}s'$ in the sum appearing in Eq.~(\ref{eq:thermal_conductivity_combined})), since perfectly degenerate vibrational modes are absent (or a negligible fraction of the total number of modes, since degeneracies are due to point-group symmetries\cite{RevModPhys_Maradudin68} and these are supposed to be absent in a realistic finite-size model of a glass).
For an ideal, infinitely large model of a glass, which can be described accurately by a calculation at $\bm{q}{=}\bm{0}$ only, one would obtain $\kappa_{\rm P}{=}0$, since at $\bm{q}{=}\bm{0}$ acoustic vibrations have zero specific heat and the time-reversal symmetry of the dynamical matrix\cite{RevModPhys_Maradudin68} implies that the group velocities for optical vibrations are zero; 
the only non-zero contribution to the total conductivity would be the coherences conductivity  $\kappa_{\rm C}$, provided some non-zero off-diagonal velocity-operator elements exists (a condition that is verified when vibrations are not Anderson-localized \cite{allen1993thermal,allen1999diffusons}). 

\begin{figure}[htbp!]
  \centering
  \includegraphics[width=\WidthFigure]{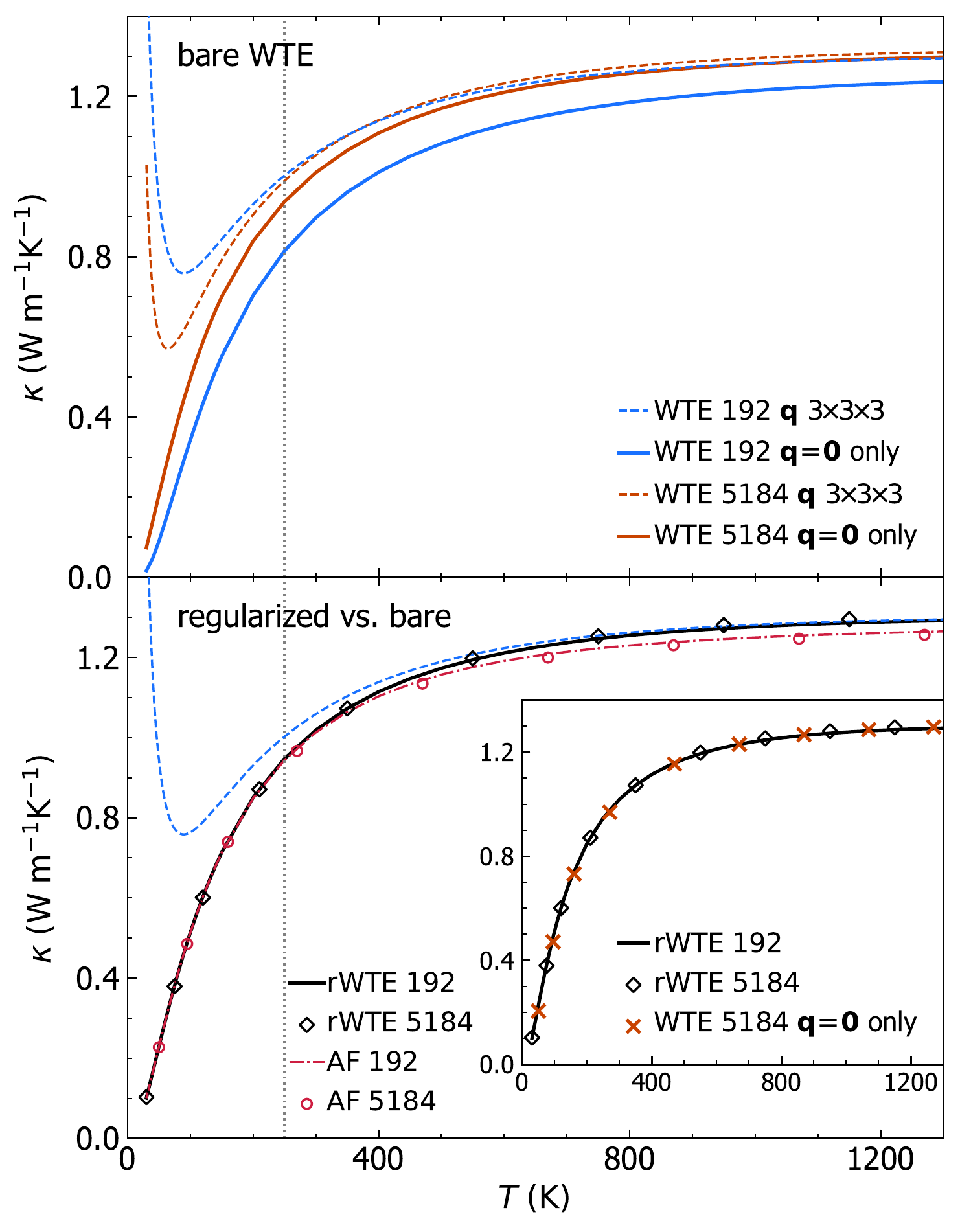}\\[-5mm]
  \caption{
{\textbf{Convergence of the anharmonic Wigner conductivity in finite-size $v$-SiO$_2$ models.} Top, bare WTE conductivity~(Eq.~(\ref{eq:thermal_conductivity_combined}) and (\ref{eq:Lorentzian})) obtained using the 192-atom (blue) or 5184-atom (orange) model, and evaluated at $\bm{q}{=}\bm{0}$  (solid lines) or on a $3{\times}3{\times}3$ $\bm{q}$ mesh to extrapolate to the bulk limit (dashed lines).
Importantly, the crystal-like divergence of the WTE conductivity computed using the $\bm{q}$-mesh is a finite-size effect, which occurs at lower temperatures for larger models. 
For $T{>}250$ K, in the 5184-atom model 
the conductivity computed at $\bm{q}{=}\bm{0}$ is practically equal to the fully $\bm{q}$-sampled conductivity; for the 192-atom model, instead, the conductivity computed at $\bm{q}{=}\bm{0}$  underestimates that limit. Importantly, for $T{>}250$ K the fully-sampled conductivity obtained from the 192-atom model is compatible with that obtained from the 5184-atom model.
Bottom panel, regularized WTE conductivity (rWTE, see text for a discussion of the regularization) for the 192- (solid black) or 5184-atom (empty diamonds) model, computed in both cases using a $3{\times}3{\times}3$ $\bm{q}$-mesh, corresponding to the full $\bm{q}$-sampling limit. The rWTE smoothly connects the correct low- and high-temperature  limits, \textit{i.e.} the fully $\bm{q}$-sampled harmonic AF conductivity (as in Fig.~\ref{fig:silica_AF_GAP}, the dashed-dotted red line is the 192-atom model, empty red circles are the 5184-atom model) and the bare WTE (dashed blue, reported from top panel), respectively.
Inset: the bare WTE conductivity evaluated at $\bm{q}{=}\bm{0}$ for the 5184-atom model is practically equal to the fully $\bm{q}$-sampled rWTE conductivities of the 192- or 5184-atom models. This demonstrates that the full $\bm{q}$-sampling limit corresponds to the bulk limit.}
  }
  \label{fig:4_effect_anharmonicity}
\end{figure}

As discussed in the previous section, a technique to extrapolate the bulk limit from finite-size models consists in relying on Fourier interpolation to sample vibrations in a $n{\times}n{\times}n$ Born von-Karman supercell of the finite periodic {reference cell}, 
while remaining aware of the limitations stemming from having a disorder length scale limited by the size of the reference cell.
Fig.~\ref{fig:linewidths_sampling}\textbf{b)} shows that such Fourier interpolation allows to greatly improve the accuracy 
of the thermodynamic predictions for the 192-atom model; specifically, the vibrational density of states (vDOS) of the 192-atom model computed using a $3{\times}3{\times}3$ $\bm{q}$-mesh is in very good agreement with the vDOS of the 5184-atom model computed at $\bm{q}{=}\bm{0}$ only.
For reference cells larger than a certain size, one expects the conductivities computed at $\bm{q}{=}\bm{0}$ only or using a $n{\times}n{\times}n$ Fourier interpolation to be practically indistinguishable\footnote{To support this expectation, we note that the sum over the modes appearing in Eq.~(\ref{eq:thermal_conductivity_combined}) contains $(3N_{at})^2$ term, out of which only $3N_{at}$ are diagonal ($s=s'$) and thus vanishing at $\bm{q}{=}\bm{0}$. This implies the difference between a conductivity calculation at $\bm{q}{=}\bm{0}$ only and one on the $n{\times}n{\times}n$ mesh goes to zero with a speed roughly proportional to the ratio between the number of diagonal elements and total number of elements, \textit{i.e.} $\frac{3N_{at}}{(3N_{at})^2}=\frac{1}{3N_{at}}$.}.
In practice, to test the accuracy of the bulk-limit extrapolations performed using the $\bm{q}$ interpolation, and to verify the aforementioned expectations, one has to perform calculations in models having different sizes (ideally in one large model that already describes the bulk limit and for which the $\bm{q}$ interpolation is therefore not needed, and in one small model where using or not the interpolation technique is expected to yield appreciable effects). 
In Fig.~\ref{fig:4_effect_anharmonicity} (upper panel) we compare the WTE conductivity~(Eq.~(\ref{eq:thermal_conductivity_combined}) and (\ref{eq:Lorentzian})) obtained using the 192-atom model (blue) or the 5184-atom model (orange) calculated either using $\bm{q}{=}\bm{0}$ only (solid lines) or using a $3{\times}3{\times}3$ Fourier interpolation (dashed lines). 
These two calculations differ only in the low-temperature limit, where they yield conductivities having opposite trends. 
The divergence of the WTE conductivity computed using the Fourier interpolation is a finite-size effect arising from the periodic boundary conditions (a reminiscence of the divergence at low temperatures of the conductivity of bulk crystals {that is cutoff in real crystals by the scattering with grains' or samples' boundaries)}; such an effect occurs at lower temperatures for larger modes; thus, it is expected to vanish in the ideal glass limit, \textit{i.e.} for $N_{\rm at}{\to} \infty$ $\kappa(T)$ in the above-the-plateau regime is expected to follow the same trend observed in the calculation at $\bm{q}{=}\bm{0}$ and in experiments (more on this later).
We note, in passing, that this reasoning also explains the opposite trend of the broadening AF conductivity curve computed  at $\bm{q}{=}\bm{0}$ only or using the Fourier interpolation  shown in Fig.~\ref{fig:harm_theory_plateau}. 
In fact, in the limit of vanishing broadening $\eta{\to}0$ only the term $s{=}s'$ determines the value of the sum in Eq.~(\ref{eq:thermal_conductivity_combined}), yielding  $\kappa{\approx}\frac{1}{\mathcal{V} N_C }\sum_{\bm{q}s}C[\omega(\bm{q})_{s}]\frac{|\!|{\tenscomp{v}(\bm{q})_{s,s}}|\!|^2}{\eta}$. 
Such a limiting expression for the conductivity diverges when the Fourier interpolation is adopted, since at $\bm{q}{\neq}\bm{0}$ the diagonal velocity-operator elements are non-zero; in contrast, in a calculation at $\bm{q}{=}\bm{0}$ only the time-reversal symmetry implies that the diagonal velocity-operator elements are zero; thus, in this case the conductivity approaches zero when the broadening approaches zero.

We note that for the 5184-atom model at temperatures higher than 100 K the WTE conductivity obtained relying on the Fourier interpolation is practically equal to the conductivity obtained evaluating the WTE conductivity at $\bm{q}{=}\bm{0}$ only. In contrast, for the 192-atom model the conductivity obtained from a calculation at $\bm{q}{=}\bm{0}$ only is always significantly different from that obtained using a computationally converged $3{\times}3{\times}3$  $\bm{q}$ mesh\footnote{Using a denser $5{\times}5{\times}5$ mesh produces practically equivalent results, implying that in the 192-atom model a $3{\times}3{\times}3$ mesh is sufficient to obtain computational convergence.}. Most importantly, for temperatures higher than 250 K the WTE conductivity obtained from the 192-atom model using the Fourier interpolation is practically indistinguishable from that obtained from the 5184-atom model, confirming that the interpolation allows to accelerate convergence in the calculation of the bulk limit.

The considerations above show that in finite-size models the WTE conductivity with Fourier interpolation is accurate up to a lower-bound temperature $T_L$ (roughly defined as the temperature at which the temperature-conductivity curve changes concavity).
Below $T_L$ finite-size effects lead to a crystal-like divergence, which emerges as a consequence of having a significant number of vibrational modes with anharmonic linewidths $\hbar\Gamma(\bm{q})_{s}$ smaller than $\hbar\Delta\omega_{\rm avg}$ (see Fig.~\ref{fig:linewidths_sampling}); from a microscopic viewpoint, this implies that couplings in the distribution~(\ref{eq:Lorentzian}) between quasi-degenerate eigenstates --- which are present in an ideal glass, see Sec.~\ref{sub:simulating_glasses_with_allen_feldman_theory_} --- are not correctly accounted for. Thus, the finite-size model fails to represent the harmonic conduction described by the distribution~(\ref{eq:Gaussian}) and accurate for an ideal glass at very low temperature\footnote{here ``very low temperature'' must be interpreted keeping into account that this work is focused on the above-the-plateau temperature regime, \textit{i.e.} $T>30 K$.}. 
This limitation can be overcome relying on the fact that in the low-temperature limit anharmonicity progressively phases out; thus, the AF model becomes increasingly more accurate and can be evaluated using the protocol discussed in the previous section.
Therefore, we introduce a regularization protocol for the WTE that allows to determine its bulk limit using Fourier interpolation, and accounts for the prescriptions needed to correctly evaluate the low-temperature harmonic limit discussed before. 
Specifically, we choose for our protocol a Voigt profile \cite{ida2000extended}  ---  a two-parameter distribution $\mathcal{F}_{[\Gamma(\bm{q})_s{+}\Gamma(\bm{q})_{s'},\eta]}$ obtained as a convolution between a Lorentzian with FWHM $\Gamma(\bm{q})_s{+}\Gamma(\bm{q})_{s'}$ and a Gaussian with variance $\eta^2\pi/2$  ---  in place of the one-parameter distribution $\mathcal{F}_{[\Gamma(\bm{q})_s{+}\Gamma(\bm{q})_{s'}]}$ appearing in Eq.~(\ref{eq:thermal_conductivity_combined}). By doing so Eq.~(\ref{eq:thermal_conductivity_combined}) reduces to the AF harmonic limit at low temperatures (where $\Delta\omega_{\rm avg}{\sim }\eta{\gg} \frac{1}{2}(\Gamma(\bm{q})_s{+}\Gamma(\bm{q})_{s'})$), and to the anharmonic WTE at intermediate and high temperatures (where $\Delta\omega_{\rm avg}{\sim} \eta{\ll} \frac{1}{2}(\Gamma(\bm{q})_s{+}\Gamma(\bm{q})_{s'})$). Hereafter the conductivity computed using the Fourier interpolation (to determine the bulk limit) and the Voigt distribution (to correctly describe the low-temperature harmonic limit) will be referred to as ``regularized WTE'' (rWTE), to distinguish it from the ``bare'' WTE (Eq.~(\ref{eq:thermal_conductivity_combined}) with the Lorentzian distribution~(\ref{eq:Lorentzian})).
As shown in the bottom panel of Fig.~\ref{fig:4_effect_anharmonicity}, the rWTE conductivity reduces to the AF harmonic conductivity for temperatures lower than $T_L$, and to the bare anharmonic WTE conductivity for temperature higher than $T_L$, smoothly connecting these two limits. 
The inset of Fig.~\ref{fig:4_effect_anharmonicity} highlights that the rWTE conductivity evaluated on a $3{\times}3{\times}3$ $\bm{q}$-mesh for the 192-atom model is 
practically indistinguishable from the rWTE conductivity of the 5184-atom model (we note in passing that for the 5184-atom model the WTE conductivity evaluated at $\bm{q}{=}\bm{0}$ only is equal to the rWTE conductivity, showing that the 5184-atom model is large enough to describe the bulk limit without relying on the $\bm{q}$ interpolation and on the regularization). Therefore, the regularization protocol greatly accelerates the convergence of the conductivity calculation, allowing to determine the bulk limit of the conductivity  of $v-$SiO$_2$ using models containing less than 200 atoms.

\section{First-principles calculations and comparison with experiments}
\begin{figure*}
  \centering
  \includegraphics[width=\textwidth]{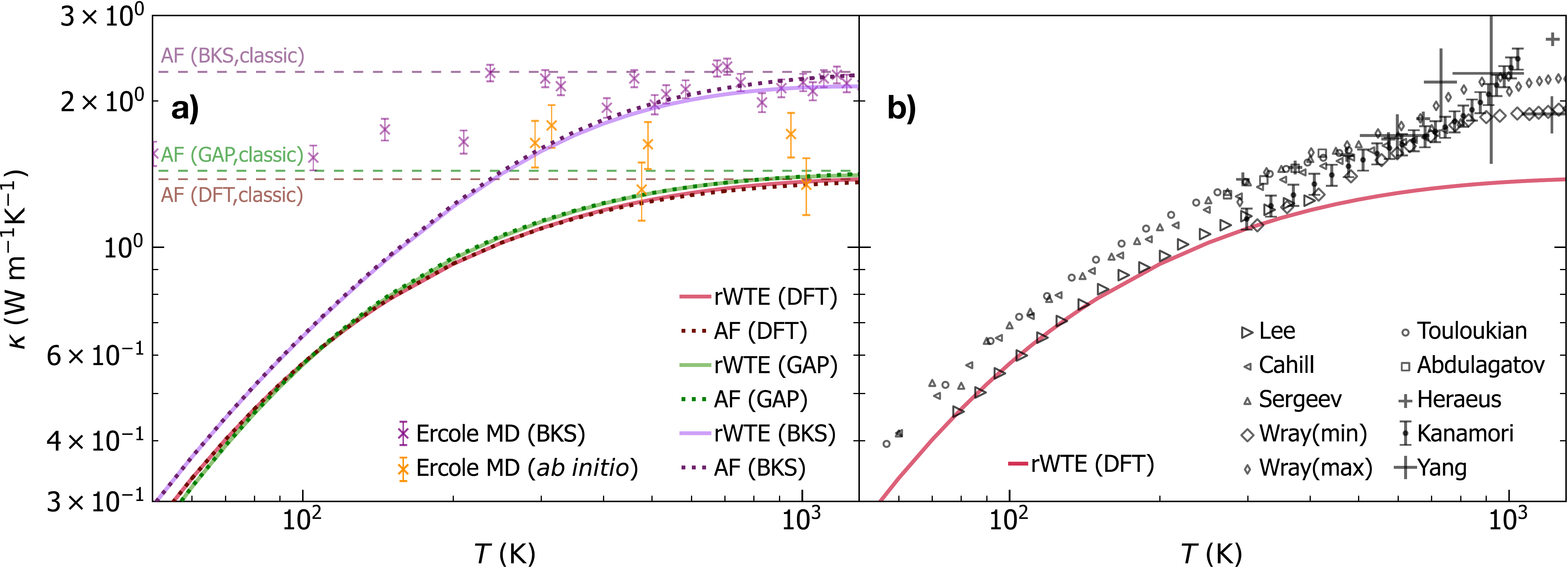}\\[-2mm]
  
  \caption{
\textbf{Thermal conductivity of $v$-SiO$_2$: theory vs experiments.} 
Panel \textbf{a)}, solid (dotted) lines are the rWTE (AF) conductivities for a 192-atom model of $v$-SiO$_2$ generated using first-principles techniques\cite{Kroll_2013} (denoted with ``192(D)'' in the text), repeated periodically $3{\times}3{\times}3$ times and then studied using directly first-principles DFT calculations (red), a GAP potential (green), or the semi-empirical BKS potential (purple).
{The dashed lines are the classical AF conductivities (represented using the same color code as before), corresponding to atomic vibrations having specific heat determined by equipartition and equal to $k_B$ for all modes \cite{PhysRevMaterials.3.085401}.
Cyan and purple scatter points are predictions from first-principles and classical (BKS-based) molecular dynamics simulations, respectively, from Ercole \textit{et al.}\cite{ercole2018ab}.
These MD calculations are governed by classical equipartition \cite{PhysRevMaterials.3.085401}, they generally agree with the corresponding classical AF conductivities (increasingly larger differences between MD and classical AF observed lowering temperature might be due to the 
decrease of the accuracy with which atomic vibrations are sampled in MD simulations when temperature is lowered).}
We also note that in the high-temperature limit, when the correct quantum Bose-Einstein statistics of vibrations yields a specific heat approaching the classical frequency-independent limit, the MD calculations are in good agreement with the corresponding (first-principles- or BKS-based) rWTE predictions. 
Panel \textbf{b)}, gray scatter points are experiments from Wray \textit{et al.} \cite{wray1959thermal}, Kanamori \textit{et al.} \cite{kanamori1968thermal}, Touloukian \textit{et al.} \cite{touloukian1971thermophysical}, Sergeev \textit{et al.} \cite{sergeev1982thermophysical}, Cahill \cite{cahill1990thermal}, Lee and Cahill (190 nm sample) \cite{lee1997heat}, Abdulagatov \textit{et al.} \cite{abdulagatov2000thermal}, Yang \textit{et al.} \cite{yang2009thermal}, and Heraeus \cite{Heraeus2010}. 
The red line represents the (most accurate) rWTE conductivity obtained from first principles, which
is in agreement with experiments in the temperature range where radiative effects are negligible ($50\lesssim T\lesssim 450$ K).
Above 500 K, our calculations depart from the increasing trend observed in experiments, showing that additional theoretical work is needed to analyze the regime where radiative effects may be relevant.
}
  \label{fig:tc}
\end{figure*}
Hitherto, no theoretical work has managed to evaluate the thermal conductivity of $v$-SiO$_2$ (and more generally of any amorphous solid) from first principles and accounting for the interplay between anharmonicity, disorder, and the Bose-Einstein statistics of atomic vibrations. 
More precisely, past works studied the thermal conductivity of $v$-SiO$_2$ using a variety of approaches, including: (i) the AF model in combination with empirical \cite{feldman1995_harm_diffusivity_SiO2} or semi-empirical 
\cite{McGaughey2009predicting} potentials; (ii) classical molecular dynamics \cite{Jund1999,McGaughey2009predicting,tian2017thermal,lv2016_locons}; (iii) first-principles molecular dynamics \cite{ercole2018ab}.
We note that the first-principles GKMD simulation by Ercole \textit{et. al.}\cite{ercole2018ab} represents a significant step forward in the field --- it is the first study that evaluates the conductivity of \mbox{$v$-SiO$_2$} from first principles --- with Ref.\cite{ercole2018ab} focusing on the high-temperature regime, where the difference between the actual quantum Bose-Einstein occupation numbers of vibrations and the classical (equipartition-determined) occupation numbers implicit in the GKMD simulations\cite{PhysRevMaterials.3.085401} is minimal.
We have seen the rWTE protocol discussed in the previous section allows to accurately determine the bulk thermal conductivity of glasses using models with a size ($\lesssim$ 200 atoms) that is within the reach of first-principles techniques; so we now employ these to compute the conductivity of $v$-SiO$_2$.

We show in Fig.~\ref{fig:tc}\textbf{a)} the bulk rWTE conductivity (solid lines) of a ``192(D)'' model of \mbox{$v$-SiO$_2$}, which contains 192 atoms and was generated relying on density-functional theory (DFT) \cite{Kroll_2013,charpentier2009first} (this model is different from the 192-atom model generated with GAP and discussed in Sec.~\ref{sec:glasses_in_periodic_boundary_conditions}; to avoid confusion we will  henceforth denote the 192-atom model discussed in Sec.~\ref{sec:glasses_in_periodic_boundary_conditions} with  ``192(G)''). 
To assess the effects of anharmonicity on the conductivity, also the harmonic AF conductivity is shown (dotted lines). 
All the parameters entering in the rWTE or AF expressions have been evaluated either from first-principles (red, 
see Methods for details), or using a state-of-the-art GAP potential\cite{erhard2022machine} (green), or using the well known semi-empirical BKS potential\cite{PhysRevLett.64.1955,carre2007amorphous} (purple)\footnote{In all these cases we used a broadening $\eta=4$ cm$^{-1}$ for the Voigt distribution, this value was determined from a convergence test analogous to that reported in the upper-left panel of Fig.~\ref{fig:harm_theory_plateau}.}.
Orange and purple scatter points are results from first-principles\footnote{Obtained using the PBE exchange-correlation functional.} GKMD and classical GKMD using the BKS potential, respectively, by Ercole \textit{et al.} \cite{ercole2018ab}; these are meant to be accurate at high temperature where the quantum Bose-Einstein vibration occupation numbers approach the classical (equipartition) occupations underlying the GKMD simulations \cite{PhysRevMaterials.3.085401}. 
Importantly, Fig.~\ref{fig:tc}\textbf{a)} shows that the conductivity of \mbox{$v$-SiO$_2$} is negligibly affected by anharmonicity. In fact, the rWTE perturbatively accounts for anharmonicity at the lowest cubic order, and yields a thermal conductivity that is (i) practically indistinguishable from that obtained employing the harmonic Allen-Feldman theory over the entire temperature range analyzed (30-1300~K); 
(ii) in good agreement at high temperature ($T {\gtrsim} 500 $~K)
with the conductivity obtained by Ercole and Baroni using GKMD \cite{ercole2018ab}, which accounts for anharmonicity exactly (in the high-temperature limit our rWTE calculations, either from first-principles or based on the BKS potential, are in very good agreement with the corresponding GKMD predictions). 
Fig.~\ref{fig:tc}\textbf{a)} reinforces the notion that the quantum Bose-Einstein statistics of vibrations plays a crucial role in determining the thermal conductivity \cite{PhysRevMaterials.3.085401} at low temperatures, since  the conductivities computed accounting for the quantum statistics (and for the correct bulk limit) increase up to saturation with temperature, while the GKMD conductivities are governed by classical equipartition and are roughly independent from temperature.
In addition, we highlight how studying the same 192(D) structure using first-principles techniques (red) or the GAP potential (green) yields very similar results, further endorsing the notion that machine-learned GAP potentials can have an accuracy comparable to that of first-principles techniques\cite{PhysRevMaterials.2.013808,deringer2021origins}. In contrast, using the semi-empirical BKS potential yields results that are significantly different from those obtained from first-principles. 
\begin{figure}[t]
\vspace*{-2mm}
  \includegraphics[width=\WidthFigure]{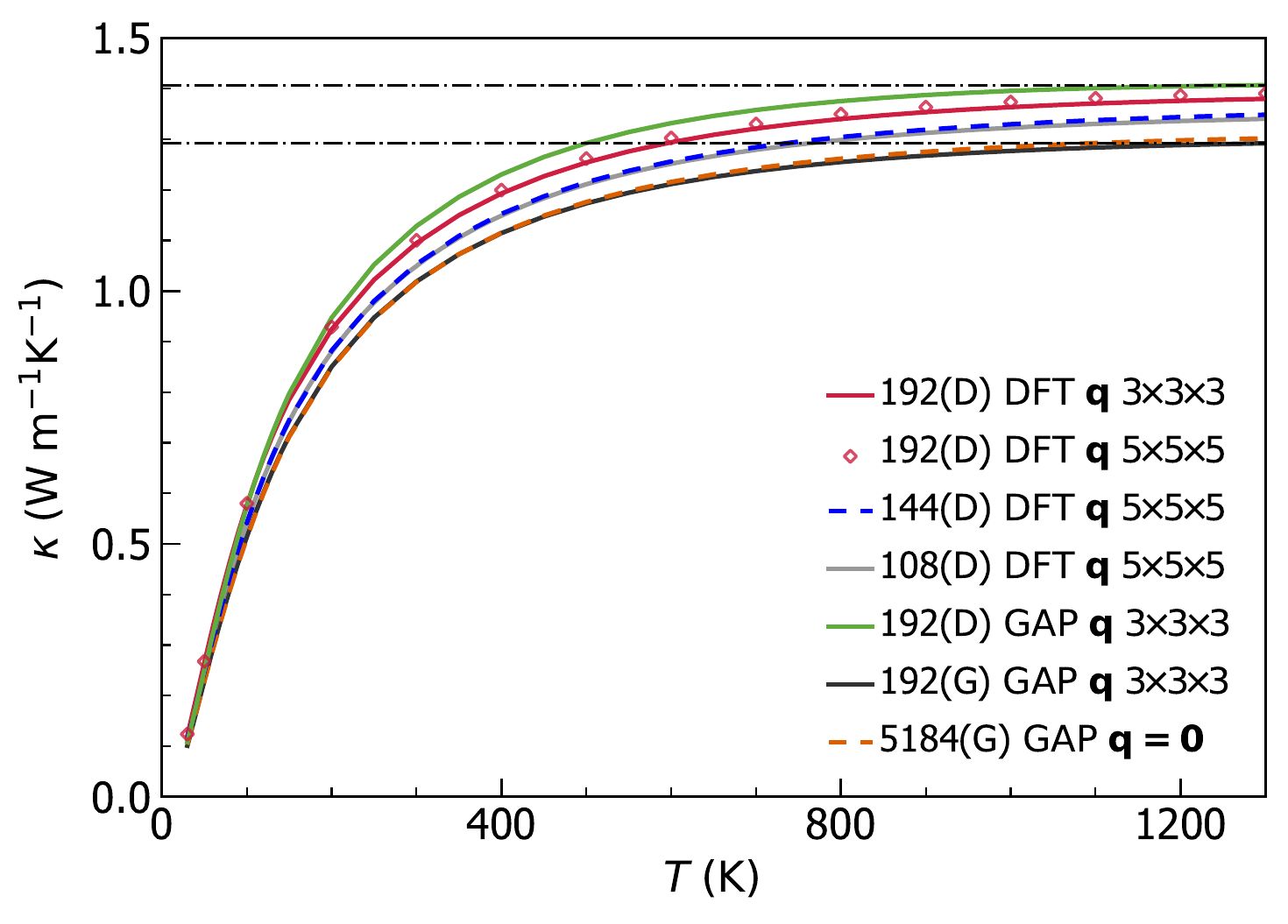}\\[-4mm]
  \caption{\hspace*{-2.mm}\textbf{rWTE} \textbf{conductivity} \textbf{of} \textbf{various} $v$-\textbf{SiO$_2$} \textbf{models}.
  Solid red is the first-principles rWTE conductivity of the 192-atom model generated relying on first-principles calculations\cite{Kroll_2013}, computed using a $3{\times}3{\times}3$ $\bm{q}$ mesh; the empty diamonds show that using a denser $5{\times}5{\times}5$ $\bm{q}$ mesh does not yield significant changes.
  The solid-gray and dashed-blue lines show the first-principles rWTE conductivity computed using a $5{\times}5{\times}5$ $\bm{q}$ mesh for a 108-\cite{charpentier2009first} and 144-atom\cite{PhysRevB.79.064202} model, respectively; both models have been generated relying on first-principles calculations. For completeness, the conductivities of the 192- (black line) and 5184-atom (dashed orange) models discussed in Fig.~\ref{fig:4_effect_anharmonicity} are reported (these are labeled with the suffix ``(G)'' to distinguish them from the structures generated relying on first-principles techniques, which are labeled with ``(D)''); the horizontal dashed-dotted lines highlight the $9\%$ difference between the rWTE conductivities of the 192(D) and 192(G) models, both described with GAP. All these atomistic models are available on the MaterialsCloud archive\cite{Materials_cloud_release}.}
  \label{fig:other_models}
\end{figure}
Importantly, Fig.~\ref{fig:tc}\textbf{b)} shows that the first-principles rWTE conductivity is in good agreement with experiments in the temperature range from 50 to $\sim$450 K; at higher temperatures the rWTE predicts a saturating trend for the conductivity (which mirrors the saturating trend of the specific heat, see Fig.~\ref{fig:spec_heat} in the Methods section), while experiments display an increasing drift. 
{ Past experimental work (e.g. Ref.~\cite{bouchut2004fused} and references therein) hinted that radiative energy transfer causes an increase of the thermal conductivity in vitreous silica at high temperature, and highlighted how accurately distinguishing the conduction and radiation contributions to heat transfer is particularly challenging\cite{yang2009thermal} (among the various sources of difficulty, Refs.~\cite{kanamori1968thermal,lee1960radiation} mentioned how the interplay between conduction and radiation depends from the size and shape of the sample used in the experiment). Thus, in Fig.~\ref{fig:tc} the discrepancy between theory and experiments at high temperature might be due to having non-negligible radiative contributions in the experiments \cite{yang2009thermal,kanamori1968thermal}, and to not accounting for these radiative contributions in our calculations.}

In Fig.~\ref{fig:tc}, the bulk limit has been computed using a $3{\times}3{\times}3$ Fourier-interpolation mesh (thus corresponding to vibrations in a system containing $192{\cdot}3^3{=}5184$ atoms); Fig.~\ref{fig:other_models} demonstrates that using this sampling achieves computational convergence, since increasing the mesh to $5{\times}5{\times}5$ yields practically indistinguishable results (empty red diamonds refer to a $5{\times}5{\times}5$ mesh, solid red line refers to a $3{\times}3{\times}3$ mesh). 
Fig.~\ref{fig:other_models} also shows that studying both the 192(D) and the 192(G) models with GAP gives high-temperature limits for their rWTE conductivities differing by about 9$\%$. 
Such a small difference 
is particularly reassuring, given the two different methods employed to produce these models (bond switching in the first case\cite{Kroll_2013,charpentier2009first} and melt-quench in the second case\cite{erhard2022machine}).
Moreover, Fig.~\ref{fig:other_models}  shows that the rWTE conductivity computed from first principles using the 192(D) model (solid red) is very similar to the conductivity computed from first-principles density-functional theory using  
models containing 108 \cite{charpentier2009first} (solid gray) or 
144 atoms \cite{PhysRevB.79.064202,Materials_cloud_structure} (dashed blue), both generated relying on first-principles calculations (therefore, we henceforth refer to these models to as ``108(D)'' and ``144(D)'', respectively).

\section{Velocity operator, anharmonicity, and trend of $\kappa(T)$} 
\label{sec:velocity_operator_and_trend}
\begin{figure*}
  \centering
\includegraphics[width=\textwidth]{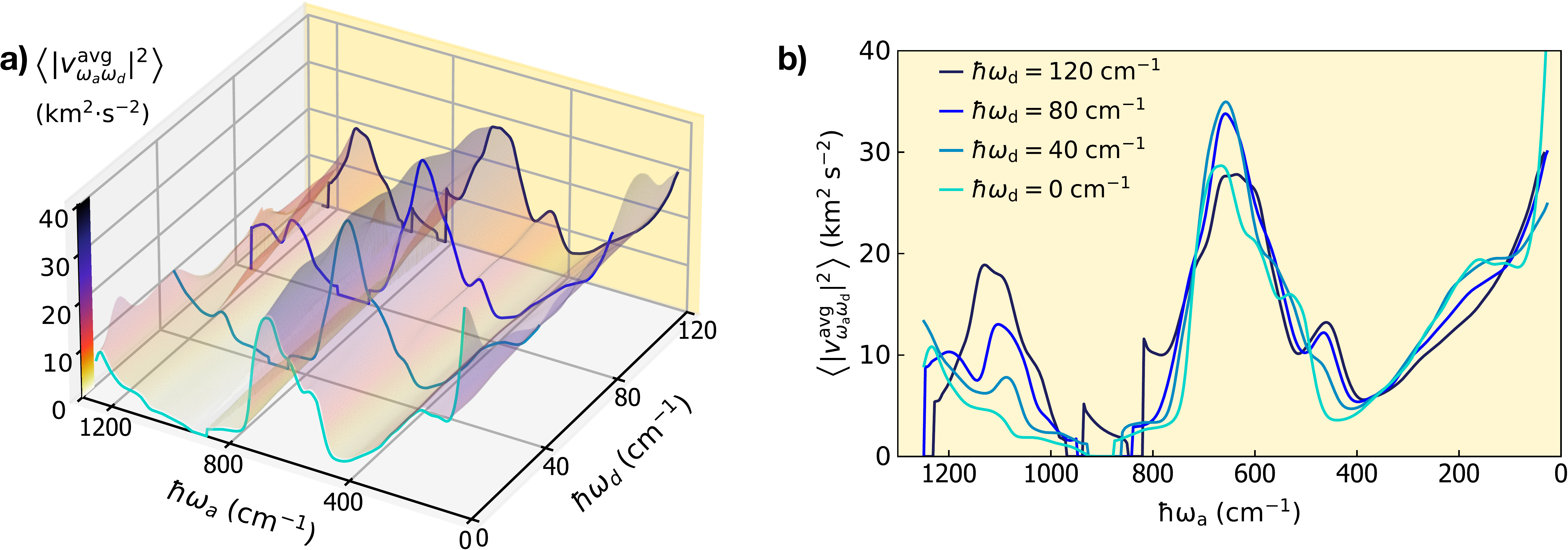}\\[-3mm]
  \caption{\textbf{Velocity operator of vitreous silica and conductivity saturation with temperature.} 
  Panel \textbf{a)}, average square modulus of the velocity-operator elements $\big<|\tenscomp{v}^{\rm avg}_{\omega_{\rm a}\omega_{\rm d}}|^2\big>$ for the 192(D) model of $v$-Si$O_2$, computed from first principles and represented as a function of the energy differences ($\hbar\omega_{\rm d}{=}\hbar(\omega(\bm{q})_{s}{-}\omega(\bm{q})_{s'})$) and averages ($\hbar\omega_{\rm a}{=}\hbar\frac{\omega(\bm{q})_{s}{+}\omega(\bm{q})_{s'}}{2}$) of the two coupled eigenstates (having wavevector $\bm{q}$ and modes $s$, $s'$; see text for details).  
  The one-dimensional projections in panel \textbf{b)} show that the elements $\big<|\tenscomp{v}^{\rm avg}_{\omega_{\rm a}\omega_{\rm d}}|^2\big>$ are almost unchanged at a given average for increasingly large energy differences. For increasingly larger temperatures, these almost-constant elements drive the saturation the rWTE conductivity (Eq.~(\ref{eq:thermal_conductivity_combined}) with the Voigt distribution discussed in Sec.~\ref{sub:Protocol_to_evaluate_the_anharmonic_Wigner_conductivity}), yielding results very close to the Allen-Feldman conductivity curve (Fig.~\ref{fig:tc}). 
  } 
  \label{fig:v_operator}
\end{figure*}
Now we want to rely on the results obtained for \mbox{$v$-SiO$_2$} to gain general insight on 
how anharmonicity affects the high-temperature trend for the $\kappa(T)$ curve in glasses.
We start by recalling that, according to the expression for the rWTE conductivity (Eq.~(\ref{eq:thermal_conductivity_combined}) with the Voigt distribution discussed in Sec.~\ref{sub:Protocol_to_evaluate_the_anharmonic_Wigner_conductivity}), at high temperatures  heat conduction in glasses is mainly determined by couplings between vibrational eigenstates, and the strengths of these couplings are given by the average square modulus of the velocity-operator elements $\frac{1}{3}\lVert\tens{v}(\bm{q})_{s,s'}\rVert^2$.
More precisely, in the high-temperature limit the specific heat is close to the classical limit and thus can be considered independent from temperature, and the linewidths are generally larger than the average energy-level spacing (Fig.~\ref{fig:linewidths_sampling}); therefore, the rWTE conductivity reduces to a Lorentzian-weighted average of velocity-operator elements. 
The weights in such average depend on the energy difference between the eigenstates coupled, and the temperature-dependent anharmonic linewidths are the  scale parameters (FWHM) of the Lorentzian distributions~(Eq.~(\ref{eq:Lorentzian})).
Increasing temperature yields larger linewidths (Fig.~\ref{fig:linewidths_sampling}), resulting  
in a broader Lorentzian distribution~(\ref{eq:Lorentzian}) that, when determining the conductivity, gives more weight to velocity-operator elements with larger energy differences.
Therefore, the variation of the elements of the velocity operator with respect to the energy differences $\hbar\omega_{\rm d}{=}\hbar(\omega(\bm{q})_{s}{-}\omega(\bm{q})_{s'})$ determines how the conductivity varies with increasing temperature\footnote{In this work the velocity operator is considered independent from anharmonicity and temperature, since we employ the standard approximation of considering the force constants to be independent from anharmonicity and temperature (for vitreous silica this approximation is realistic, as discussed in the Methods).}: matrix elements increasing or decreasing with $\omega_d$ imply a conductivity increasing or decreasing with temperature. 
From this reasoning and from the saturating trend of the temperature-conductivity curve for vitreous silica shown in Fig.~\ref{fig:tc}, we expect the velocity-operator elements for \mbox{$v$-SiO$_2$} to be approximately constant with respect to $\omega_d$.
To verify this prediction, we plot in Fig.~\ref{fig:v_operator}\textbf{a)} the velocity-operator elements for $v$-SiO$_2$ (we show the velocity operator for the 192(D) model and computed from first principles; the other models yield practically indistinguishable results when analyzed from first principles or using GAP) as a function of the energy difference $\hbar\omega_{\rm d}$ and of the energy average $\hbar\omega_{\rm a}{=}\hbar({\omega(\bm{q})_{s}{+}\omega(\bm{q})_{s'}})/{2}$:
\begin{equation}
\begin{split}
  &\big<|\tenscomp{v}^{\rm avg}_{\omega_{\rm a}\omega_{\rm d}}|^2\big>=[\mathcal{G}(\omega_{\rm a},\omega_{\rm d})]^{-1}\frac{1}{\mathcal{V}N_{\rm c}}{\sum_{\bm{q},s,s'}} \frac{\rVert\tens{v}(\bm{q})_{s,s'}\rVert^2}{3}\\
  &\hspace*{1cm}{\times} 
 \delta\left(\omega_{\rm d}{-}(\omega(\bm{q})_s{-}\omega(\bm{q})_{s'})\right)\delta\Big(\omega_{\rm a}{-}\frac{\omega(\bm{q})_s{+}\omega(\bm{q})_{s'}}{2} \Big)\;;
 \raisetag{15mm}
  \label{eq:v_operator_omega_a_omega_d}
\end{split}
\end{equation}
$\mathcal{G}(\omega_{\rm a},\omega_{\rm d})$ is a density of states that serves as normalization (see Methods for details).  
The one-dimensional projections in Fig.~\ref{fig:v_operator}\textbf{b)} show that these velocity-operator elements are almost constant at varying $\hbar\omega_{\rm d}$ for all values of $\hbar\omega_{\rm a}$. 
These findings validate the reasoning above: the variation of the velocity-operator elements $\frac{1}{3}\lVert\tens{v}(\bm{q})_{s,s'}\rVert^2$ with respect to the energy differences $\hbar\omega_{\rm d}$ determines how the conductivity varies with increasing temperature, implying that a saturating temperature-conductivity curve is obtained when most of the velocity-operator elements do not vary with $\hbar\omega_{\rm d}$.
The reasoning above also explains the small or negligible effects of anharmonicity on the conductivity of $v$-SiO$_2$ discussed before, since having a velocity operator displaying a negligible dependence on $\hbar\omega_{\rm d}$ implies that the rWTE conductivity does not vary appreciably when the linewidths (broadening of the Lorentzian distribution) vary, and therefore it does not significantly differ from the AF conductivity (computed using a constant broadening $\eta$).

\section{Thermal diffusivity}
In order to gain further insight on the microscopic mechanisms underlying conduction it is useful to resolve how each vibrational mode contributes to transport; \textit{i.e.}, the quantity of heat that it carries and the rate at which it diffuses. 
It is possible to extract the contribution of a single vibration to thermal transport by factorizing the single-vibration specific heat $C(\bm{q})_s$ in the regularized rWTE conductivity~(\ref{eq:thermal_conductivity_combined}) recasting it as $\kappa=\frac{1}{\mathcal{V}N_{\rm c}}\sum_{\bm{q}s} C(\bm{q})_s D(\bm{q})_s$, with $D(\bm{q})_s$ being the ``anharmonic thermal diffusivity''. The expression for $D(\bm{q})_s$ is determined by such factorization and by the requirement that in the coherences' coupling between two vibrations $(\bm{q})_{s}$ and $(\bm{q})_{s'}$ each  contributes to the coupling with a weight equal to the relative specific heat \cite{simoncelli2021Wigner} (e.g. for vibration $(\bm{q})_{s}$ the weight is $\tfrac{C(\bm{q})_s}{C(\bm{q})_s+C(\bm{q})_{s'}}$, and correspondingly for vibration $(\bm{q})_{s'}$ the weight is $\tfrac{C(\bm{q})_{s'}}{C(\bm{q})_s+C(\bm{q})_{s'}}$):
\begin{equation}
\begin{split}
D(\bm{q})_s=&{\sum_{s'}}
\frac{\omega(\bm{q})_s{+}\omega(\bm{q})_{s'\!} }{{2[C(\bm{q})_s{+}C(\bm{q})_{s'\!}] }}
\!\left[\!\frac{C(\bm{q})_s}{\omega(\bm{q})_s}{+}\frac{C(\bm{q})_{s'}}{\omega(\bm{q})_{s'}}\!\right]\!\!
\frac{\lVert\tens{v}(\bm{q})_{s,s'}\lVert^2\!}{3}\\
&\times\pi\mathcal{F}_{[\Gamma(\bm{q})_s{+}\Gamma(\bm{q})_{s'},\eta]}(\omega(\bm{q})_s-\omega(\bm{q})_{s'})\;.
\label{eq:diffusivity_q_s}
\end{split}
\raisetag{5mm}
\end{equation}
The goal of this decomposition is to resolve the rate at which the heat carried by a vibration with wavevector $\bm{q}$ and mode $s$ diffuses. 
We note that Eq.~(\ref{eq:diffusivity_q_s}) accounts for the effects of anharmonicity on vibrations' diffusion by means of the linewidths, depends on temperature through both the specific heat and the linewidths, and applies to both glasses and crystals --- in the former case the wavevector $\bm{q}$ is just a label without direct physical meaning (we will discuss later that for glasses the diffusivity has to be represented as a function of frequency to be well defined), while in the latter case such an expression is accurate if and only if the SMA is accurate and $\eta{=}0$ is used\footnote{This last condition implies that the Voigt distribution analytically reduces to the Lorentzian distribution~(\ref{eq:Lorentzian})}. It is worth mentioning that in the case of simple crystals, characterized by $\kappa_P{\gg}\kappa_C$, the term $s'{=}s$ in Eq.~(\ref{eq:diffusivity_q_s})
yields the well known expression obtained from Peierls's theory, which interprets the diffusivity (averaged over the three Cartesian directions) as $D(\bm{q})_s=\frac{1}{3}|\!|\tens{v}(\bm{q})_{s,s}|\!|^2\tau(\bm{q})_s$, where $\tens{v}(\bm{q})_{s,s}$ is the free propagation velocity of the particle-like heat carrier with wavevector $\bm{q}$ and mode $s$, and $\tau(\bm{q})_s$ is the inter-collision time.
In finite-size models of glasses it is most informative to represent the diffusivity as a function of frequency, first because  
the specific heat of a vibration depends on its frequency $\omega$ ($C(\bm{q})_s{=}C(\omega(\bm{q})_s)$, see Eq.~(\ref{eq:quantum_specific_heat_A})), and second because
the vibrational frequencies determine measurable quantities such as the vibrational density of states (in contrast, as mentioned before, in finite-size models of glasses quantities such as the wavevectors $\bm{q}$ span a BZ that depends on the model and are used only as a mathematical tool in the determination of the bulk limit).
Thus, we represent the thermal diffusivity as a function of frequency with $D(\omega,T){=}[g(\omega){\mathcal{V}N_{\rm c}}]^{-1}\sum_{\bm{q},s} D(\bm{q})_s \delta(\omega{-}\omega(\bm{q})_s)$ (here $g(\omega){=}({\mathcal{V}N_{\rm c}})^{-1}\sum_{\bm{q},s}\delta(\omega{-}\omega(\bm{q})_s)$ is the vibrational density of states (vDOS), which can be considered independent from temperature, as also shown in Fig.~\ref{fig:spec_heat} in the Methods; the Dirac $\delta$ is broadened with a Gaussian distribution having a broadening determined from the convergence test discussed in Sec.~\ref{sub:simulating_glasses_with_allen_feldman_theory_}). 
In the low-temperature and infinite-reference-cell  limit, Eq.~(\ref{eq:diffusivity_q_s})  reduces to the temperature-independent harmonic diffusivity introduced by Allen and Feldman \cite{allen1989thermal} (this follows from the properties of the Voigt distribution $\mathcal{F}_{[\Gamma(\bm{q})_s{+}\Gamma(\bm{q})_{s'},\eta]}$ discussed in Sec.~\ref{sub:Protocol_to_evaluate_the_anharmonic_Wigner_conductivity}); in the following the dependence from temperature will be shown explicitly for the sake of clarity. 
In this frequency-dependent representation 
the conductivity reads
\begin{equation}
  \kappa(T)=\int_{0}^{\infty} g(\omega) C(\omega,T) D(\omega,T) d\omega\;,
  \label{eq:kappa_diff}
\end{equation}
and intuitively allows to resolve the contribution of vibrations with frequency $\omega$ to heat transport through their density of states $g(\omega)$, the heat carried $C(\omega,T)$, and the diffusion rate $D(\omega,T)$.
\begin{figure}[htbp!]
  \centering
\begin{overpic}[width=\WidthFigure]{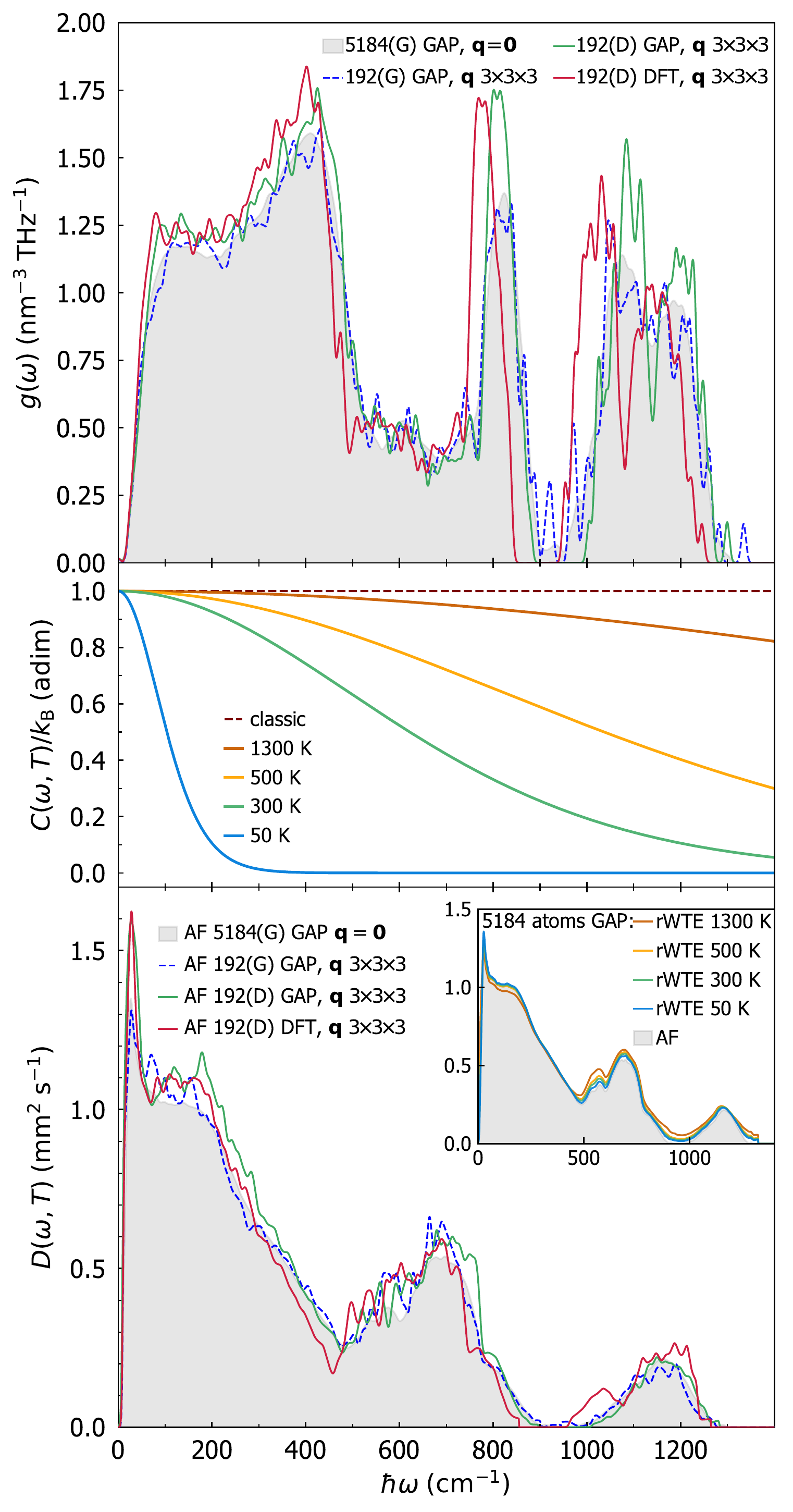}
  \put(-2,96){\textbf{a)}}
  \put(-2,61){\textbf{b)}}
  \put(-2,40){\textbf{c)}}
 \end{overpic}\\[-5mm]
  \caption{
\textbf{Vibrational DOS, specific heat, and diffusivity of vitreous silica.}
   \textbf{a)} Vibrational density of states of vitreous silica. Red and green are the vDOS for the 192(D)\cite{Kroll_2013} model (192-atom model generated from first-principles DFT simulations), obtained evaluating the frequencies of such model either from first principles or from the GAP potential, and using a $3{\times}3{\times}3$ $\bm{q}$ mesh.
   The dashed blue line and gray area are the vDOS of the GAP-generated\cite{erhard2022machine} 192- and 5184-atom models,
   computed using GAP as in Fig.~\ref{fig:4_effect_anharmonicity} (\textit{i.e.} relying on a $3{\times}3{\times}3$ $\bm{q}$ mesh in the former case, and evaluating the diffusivity at $\bm{q}{=}\bm{0}$ only in the latter case).
   \textbf{b)}~Quantum harmonic specific heat as a function of frequency and temperature, obtained evaluating the expression~(\ref{eq:quantum_specific_heat_A}) (solid lines).
   The dashed line represents the classical specific heat (obtained from the quantum specific heat in the infinite-temperature limit). 
   \textbf{c)} Harmonic Allen-Feldman thermal diffusivity computed from Eq.~(\ref{eq:diffusivity_q_s}) using $\Gamma(\bm{q})_s{=}0\;\forall\bm{q},s$ and with the models discusses in panel \textbf{a)} (same parameters and color code); the inset shows that in $v$-SiO$_2$ (5184(G) model) the AF diffusivity is very similar to the anharmonic diffusivity (full Eq.~(\ref{eq:diffusivity_q_s})).
   \vspace*{-5mm}
  }
  \label{fig:DOS_diff_spec_heat}
\end{figure} 
We report in Fig.~\ref{fig:DOS_diff_spec_heat} all these quantities.
In panel \textbf{a)} we show the vDOS for the 192(D) \mbox{$v$-SiO$_2$} model computed using first principles calculations (red) or the GAP potential (green). The main difference between these calculations is that the GAP potential slightly  stretches the high-frequency part of the spectrum towards higher values; the effects on the thermal conductivity of such a slight stretch of vibrational energies is barely appreciable, as shown in Fig.~\ref{fig:tc}. 
The dashed-blue line and the gray area are the vDOS for the 192(G) and 5184(G) models, respectively;
 both these vDOS are obtained using GAP, the former is computed using a $3{\times}3{\times}3$ $\bm{q}$ interpolation mesh, while the latter is computed at $\bm{q}{=}\bm{0}$ only. The similarity between these two curves further supports the usage of the $\bm{q}$ interpolation technique to sample more accurately the vibrations in a glass.
{Fig.~\ref{fig:DOS_diff_spec_heat}\textbf{a)} 
 shows that the vDOS is slightly affected by the method used to generated the $v$-SiO$_2$ model, with the 192(G) and 5184(G) models generated using the melt-quench method \cite{erhard2022machine} leading to vDOS slightly different from the 192(D) model, which was generated using the bond-switching technique\cite{charpentier2009first,Kroll_2013}. }
In panel \textbf{b)} of Fig.~\ref{fig:DOS_diff_spec_heat} we report the quantum harmonic specific heat as a function of frequency, showing how increasing temperature populates vibrational modes of increasingly larger frequency that consequently contribute to transport. 
We highlight how, for all the temperatures considered, the quantum specific heat differs from the constant classical (equipartition) limit (obtained letting $T{\to}\infty$ in Eq.~(\ref{eq:quantum_specific_heat_A})); this reiterates the role of quantum statistics of vibrations in thermal transport \cite{PhysRevMaterials.3.085401}.
Panel~\textbf{c)} shows the harmonic AF diffusivity (Eq.~(\ref{eq:diffusivity_q_s}) using $\Gamma(\bm{q})_s{=}0\;\forall\bm{q},s$) for the models of \mbox{$v$-SiO$_2$} discussed in panel~\textbf{a)} (same parameters and color code are used). Considerations analogous to those for the vDOS in panel~\textbf{a)} hold: the method used to generated the $v$-SiO$_2$ model has a small but appreciable effect on the diffusivity, with the 192(D) model generated using the bond-switching method having a larger diffusivity at low frequency. 
Overall, the combined variations of vDOS and diffusivity due to the technique used to generate the model, or the approach (first-principles calculations or GAP) used to evaluate frequencies and velocity operators, yield differences on the thermal conductivity within 9$\%$, as shown before in Fig.~\ref{fig:other_models}.
The inset of panel \textbf{c)} shows that the anharmonic diffusivity (colored lines, computed from Eq.~(\ref{eq:diffusivity_q_s})) changes very little with temperature, and is practically very similar to the temperature-independent AF diffusivity (gray area, obtained using Eq.~(\ref{eq:diffusivity_q_s}) with $\Gamma(\bm{q})_s{=}0\;\forall\bm{q},s$ and $\eta$ determined from the convergence test detailed in Fig.~\ref{fig:harm_theory_plateau}); the inset shows calculations with GAP on the 5184(G) model, analogous considerations hold for the other models studied from first-principles or using GAP.

\section{Conclusions}
We have discussed a computational protocol that allows to determine from finite-size models of glasses containing less than 200 atoms --- thus within the reach of standard first-principles approaches --- the bulk limit of the harmonic Allen-Feldman conductivity\cite{allen1989thermal,allen1993thermal}, as well as of the anharmonic Wigner conductivity\cite{simoncelli2019unified,simoncelli2021Wigner}. 
To determine the bulk limit of the harmonic AF conductivity the protocol employs Fourier interpolation to improve the sampling of the vibrational spectrum of the glass model, and it numerically represents the Dirac $\delta$ appearing in the AF conductivity with a light-tailed Gaussian broadening larger than the average vibrational energy-level spacing (in place of the originally proposed fat-tailed Lorentzian broadening\cite{allen1989thermal,allen1993thermal}). 
To evaluate the bulk limit of the  Wigner conductivity, the protocol uses a Voigt profile --- a two-parameter distribution obtained as a convolution between the Gaussian used in the AF calculation, and the Lorentzian with FWHM determined by the linewidths appearing in the Wigner conductivity. The Voigt profile ensures that 
the linewidths affect the conductivity only when they are larger than the broadening used in the AF calculation; this allows to retain in finite-size models of glasses the physical property that heat transfer via a wave-like tunneling between neighboring (quasi-degenerate) vibrational eigenstates can occur even in the limit of vanishing anharmonicity, provided these eigenstates are not Anderson-localized, \textit{i.e.} that they are coupled by non-zero velocity-operator elements.

The protocol has been validated on the paradigmatic glass \mbox{$v$-SiO$_2$}, using a state-of-the-art GAP potential and atomistic models containing 192 or 5184 atoms\cite{erhard2022machine} to compute vibrational states;  we have shown that employing the protocol on a 192-atom model allows to obtain harmonic (AF) and anharmonic (rWTE) conductivities in perfect agreement with those of this much larger, 5184-atom model generated using the same technique.

After validation, we have used the protocol to predict
the AF and rWTE conductivities of \mbox{$v$-SiO$_2$} fully from first-principles.
We have shown that anharmonicity does not significantly affect the conductivity of \mbox{$v$-SiO$_2$}, even at high temperatures, since the AF conductivity is very similar to the rWTE conductivity over the entire temperature range analyzed ($30{<}T{<}1300$ K).
We have supported this finding by showing that the rWTE conductivity, which accounts for anharmonicity at the lowest (third) perturbative order \cite{simoncelli2021Wigner}, is compatible at high temperatures with the conductivity obtained from first-principles GKMD from Ercole \textit{et. al.}\cite{ercole2018ab} (we recall that GKMD simulations are accurate at high temperature, where the quantum specific heat approaches the classical limit and anharmonic effects are maximized).
Our calculations are in agreement with experiments in the temperature range $50{\lesssim} T {\lesssim} 450$ K, but do not describe the surge of the conductivity observed at higher temperatures.
Future work will aim at understanding if such a discrepancy can be related, as it seems likely, to radiative effects\cite{bouchut2004fused}.

The results obtained for \mbox{$v$-SiO$_2$} have allowed us to gain general insights on how anharmonicity affects the thermal conductivity of glasses at high temperature. Specifically, we have shown that the high-temperature trend of the conductivity is determined by how off-diagonal velocity-operator elements  ---  which couple pairs of vibrational eigenstates $(\bm{q})_s$ and $(\bm{q})_{s'}$, allowing tunnelling between them  ---  vary as a function of the energy difference $\hbar\omega_d{=}\omega(\bm{q})_{s}{-}\omega(\bm{q})_{s'}$. Velocity-operator elements increasing (decreasing) with $\hbar\omega_d$ drive a conductivity increase (decrease) with temperature. In the specific case of \mbox{$v$-SiO$_2$}, the saturating trend of the conductivity derives from velocity-operator elements that are constant with respect to the energy difference between the eigenstates coupled.

Finally, we have interpreted heat conduction in terms of frequency-resolved thermal diffusivities for vibrations, showing that the harmonic Allen-Feldman diffusivity characterizes accurately thermal transport in vitreous silica at the microscopic level.
{This work paves the way to study the thermal conductivity of glasses from first principles, and will be particularly relevant to investigate the thermal properties of amorphous materials for which developing quantum-accurate interatomic potentials is particularly challenging or unpractical.}

\section*{Acknowledgements} 
\label{sec:section_name}
We thank G. Csányi and V. L. Deringer for useful discussions. 
N. M. acknowledges funding from the Swiss National Science Foundation under the Sinergia grant no. 189924.
M. S. acknowledges support from Gonville and Caius College, and from the SNSF project P500PT\_203178.
Part of the calculations presented in this work have been performed using computational resources provided by the Cambridge Tier-2 system operated by the University of Cambridge Research Computing Service (\url{www.hpc.cam.ac.uk}) funded by EPSRC Tier-2 capital grant EP/T022159/1.

\newpage

\vspace*{22cm}
\appendix

\begin{figure*}
  \centering
  \includegraphics[width=\textwidth]{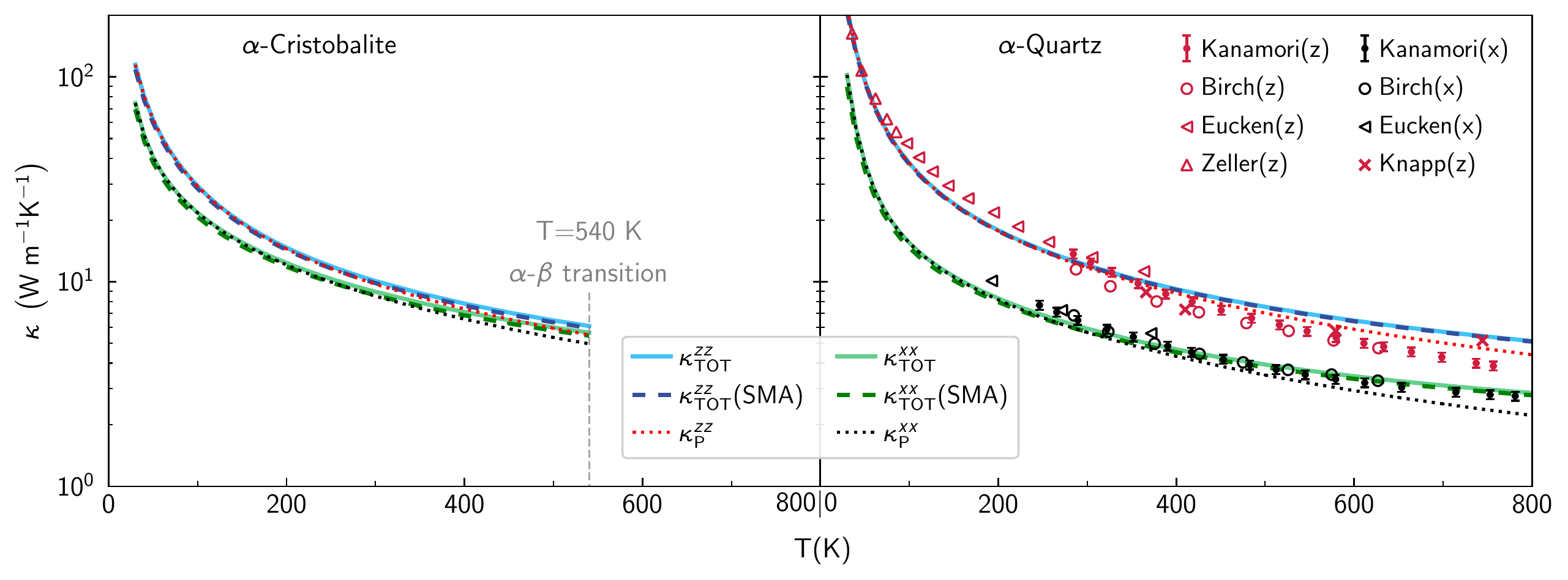}
  \caption{\textbf{Thermal conductivity of crystalline $\alpha$-cristobalite and $\alpha$-quartz.} 
  For these materials the total bulk conductivity resulting from the full solution of the WTE (solid lines) is practically indistinguishable from the bulk conductivity computed relying on the SMA approximation (dashed lines). 
  The $\alpha$-$\beta$ transition temperature for cristobalite is around 540 K \cite{Cristobalite_alpha_beta_transition_temperature}, below this temperature both $\alpha$-cristobalite and $\alpha$-quartz are to a good approximation simple crystals, since their total conductivity is predominantly determined by the particle-like contribution ($\kappa_{_{\rm P}}$, dotted lines).    
  Scatter points are experiments by Eucken \cite{eucken1911}, Birch \textit{et al.} \cite{birch1940thermal}, Knapp \cite{Knapp1943}, Zeller \textit{et al.} \cite{Zeller_Pohl_1971} and Kanamori \textit{et al.} \cite{kanamori1968thermal}. 
  }
  \label{fig:conductivity_quartz_cristobalite}
\end{figure*}
\section{Accuracy of the SMA approximation in glasses} 
\label{sec:accuracy_of_the_sma_approximation_in_glasses}

Glasses feature a low thermal conductivity, originating from strong scattering of vibrations due to anharmonicity or disorder.
Past works\cite{li2014shengbte,lindsay_first_2016,fugallo2013ab,simoncelli2021Wigner} have shown that as scattering becomes stronger, the relaxation to equilibrium of the vibrational excitations  becomes faster, and therefore the SMA --- which 
approximates the collision operator as in Eq.~\ref{eq:Lorentzian} (see Ref.\cite{simoncelli2021Wigner}) --- becomes more accurate.
These considerations already suggest that the SMA approximation is accurate in vitreous silica.
To further support these expectations, we investigated numerically the accuracy of the SMA approximation in the crystalline silica polymorphs $\alpha$-cristobalite and $\alpha$-quartz. These crystals have a conductivity much larger than vitreous silica, implying a weaker scattering from disorder or anharmonicity; consequently in $\alpha$-cristobalite and $\alpha$-quartz the SMA is expected\cite{li2014shengbte,lindsay_first_2016,fugallo2013ab} to be less accurate than in vitreous silica.
Fig.~\ref{fig:conductivity_quartz_cristobalite} shows that for $\alpha$-cristobalite and $\alpha$-quartz solving the WTE in full or employing the SMA approximation yields conductivities practically indistinguishable; this demonstrates that the SMA is accurate in $\alpha$-cristobalite and $\alpha$-quartz, and also suggests that the SMA is even more accurate in vitreous silica.

\section{Effects of disorder and temperature on the linewidths} 
\label{sec:effects_of_disorder_and_temperature_on_the_linewidths}

\begin{figure*}
  \includegraphics[width=\textwidth]{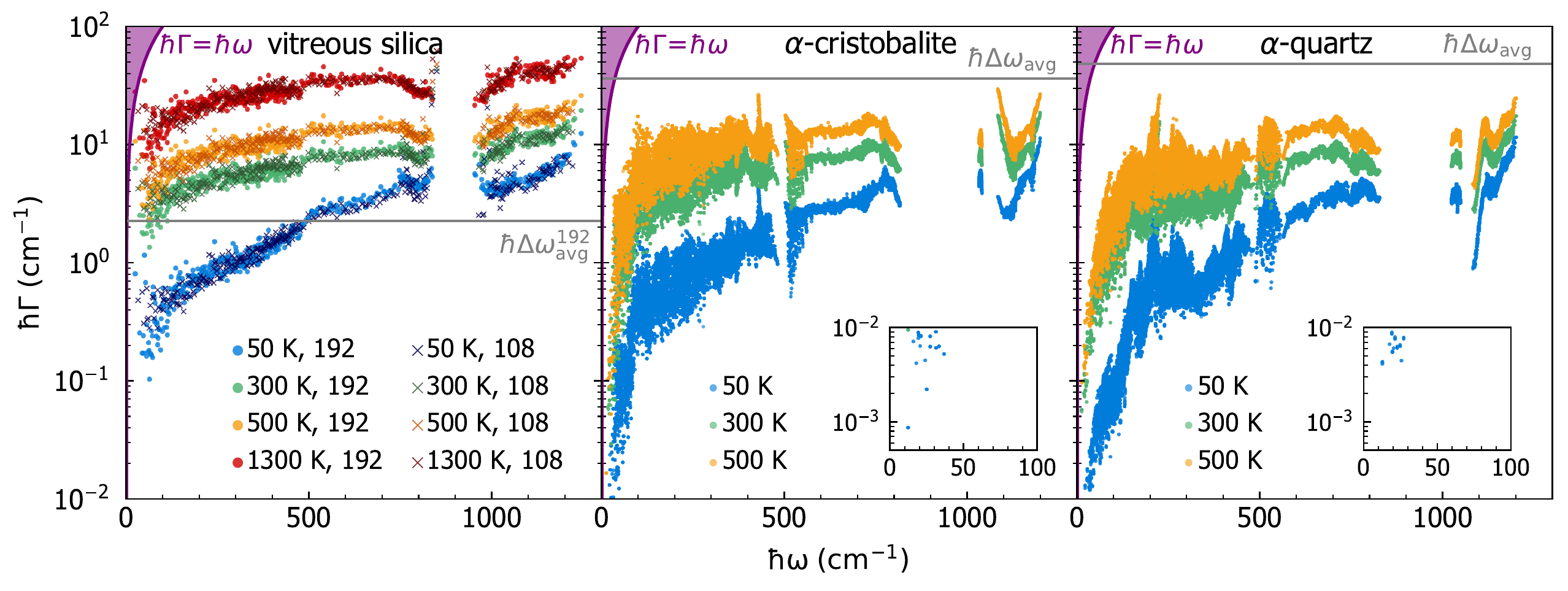}\\[-3mm]
     \caption{\textbf{Effect of disorder and temperature on the linewidths of silica polymorphs.} Increasing temperature yields an increase of the linewidths that account for third-order anharmonicity \cite{paulatto2013anharmonic,li2014shengbte} and isotopic scattering \cite{tamura1983isotope} in both vitreous and crystalline materials: the left panel shows the 192(D) and 108(D) vitreous-silica model (both computed from first-principles and at $\bm{q}{=}\bm{0}$ only), the central panel shows $\alpha$-cristobalite, and the right panel shows $\alpha$-quartz. 
    The temperatures of 50, 300, and 500 K at which the linewidths distributions are reported are chosen to span the temperature range over which all these materials are stable. The 1300 K distribution shows the behavior at high temperature, and is reported only for vitreous silica since $\alpha$-cristobalite and $\alpha$-quartz are not stable at this temperature.
    The insets in panel \textbf{b)} and \textbf{c)} show the linewidths at low vibrational energies for $\alpha$-cristobalite and $\alpha$-quartz, respectively. 
    The purple area shows the overdamped regime characterized by $\hbar\Gamma>\hbar\omega$; the lack of linewidths in the purple region shows that there are no overdamped vibrations and thus the Wigner formulation can be employed \cite{simoncelli2021Wigner,Caldarelli_2022}. The gray lines represent the average spacing between the vibrational energy levels Eq.~(\ref{eq:average_spacing}) (in panel \textbf{a)} only the average energy-level spacing for the 192-atom model is reported). 
     }
    \label{fig:lw_compare}
  \end{figure*}

In Fig.~\ref{fig:lw_compare}\textbf{a)} we highlight how the frequency-linewidth distributions for models of vitreous silica having different size (192 or 108 atoms) are overlapping, suggesting that the linewidths of these models are not significantly affected by finite-size effects.
In panel \textbf{b)} and \textbf{c)} we show the linewidths of the crystalline polymorphs  $\alpha$-cristobalite (containing 12 atoms per primitive cell) $\alpha$-quartz (9 atoms per primitive cell). Clearly, the variation of the anharmonic linewidths is mainly due to temperature, since at fixed temperature the frequency-linewidth distributions of amorphous and crystalline silica polymorphs have a similar magnitude. 
We also note that the anharmonic linewidths computed with GAP for the 192(G) model (Fig.~\ref{fig:linewidths_sampling}) are very similar to the anharmonic linewidths computed from first-principles for the 192(D) and 108(D) models (Fig.~\ref{fig:lw_compare}\textbf{a)}).

\newpage
\section{Accounting for anharmonicity at a reduced computational cost} 
\label{sec:accounting_for_anharmonicity_at_a_reduced_computational_cost}
In this section we discuss the details of the computation of the analytical function $\Gamma_a[\omega]$, used to approximatively determine the linewidths as a function of frequency discussed in Fig.~\ref{fig:linewidths_sampling}.
The analytical function $\Gamma_a[\omega]$ is determined as
\begin{equation}
  \Gamma_a[\omega]=\frac{1}{\sqrt{\frac{1}{(\Gamma_1[\omega])^2}+\frac{1}{(\Gamma_2[\omega])^2}}},
  \label{eq:approx_analytical_f}
\end{equation}
where $\Gamma_1[\omega]$ and $\Gamma_2[\omega]$ are defined as
\begin{equation}
\begin{split}
    &\Gamma_1[\omega]{=}\frac{\sum\limits_{\bm{q}=\bm{0},s}\frac{1}{\sqrt{2\pi\sigma^2}}\exp\Big[-\frac{\hbar^2(\omega(\bm{q})_s-\omega)^2}{2\sigma^2}\Big]}{\sum\limits_{\bm{q}=\bm{0},s}\tau(\bm{q})_s\frac{1}{\sqrt{2\pi\sigma^2}}\exp\Big[-\frac{\hbar^2(\omega(\bm{q})_s-\omega)^2}{2\sigma^2}\Big]},\\
    &\Gamma_2[\omega]{=}p\cdot \omega^2,\\
    &p=\frac{\sum\limits_{\bm{q}=\bm{0},s}\int_{\omega_{\rm o}}^{2\omega_{\rm o}}d\omega_c \frac{\Gamma(\bm{q})_s}{\omega^2(\bm{q})_s}
    \frac{1}{\sqrt{2\pi\sigma^2}}\exp\!\Big[{-}\tfrac{\hbar^2(\omega(\bm{q})_s-\omega_c)^2}{2\sigma^2}\Big] }{\sum\limits_{\bm{q}=\bm{0},s}\int_{\omega_{\rm o}}^{2\omega_{\rm o}}d\omega_c 
    \frac{1}{\sqrt{2\pi\sigma^2}}\exp\!\Big[{-}\tfrac{\hbar^2(\omega(\bm{q})_s-\omega_c)^2}{2\sigma^2}\Big]}.
\end{split}
\label{eq:approx_analytical_f2}
\end{equation}
$\omega_{\rm o}$ is the smallest non-zero frequency at $\bm{q}=\bm{0}$ and $\sigma{=}15$ cm$^{-1}$ is a broadening chosen sufficiently large 
to ensure that the linewidths are averaged in a smooth way.
The functional form of the approximated function $\Gamma_a[\omega]$ is inspired by past work \cite{garg2011thermal_PhD,PhysRevB.105.134202}, 
and the specific expressions~(\ref{eq:approx_analytical_f},\ref{eq:approx_analytical_f2}) to determine it have been devised and validated relying on exact calculations performed on the 108(D) $v$-SiO$_2$ model.
\begin{figure}[htbp!]
  \centering
  \includegraphics[width=\WidthFigure]{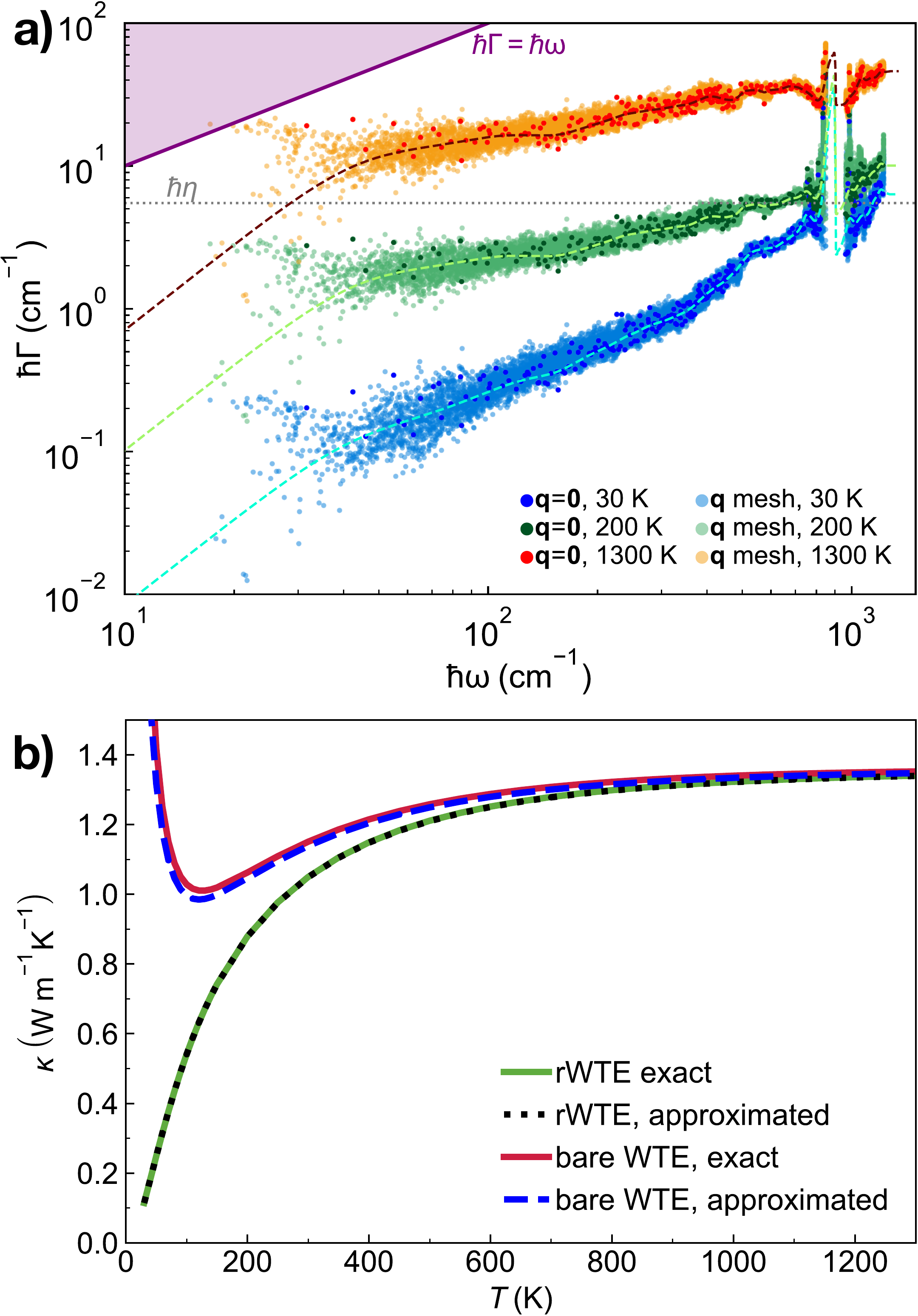}
  \caption{\textbf{Effect of $\bm{q}$-interpolation on the linewidths and approximation to reduce the computational cost.} Panel~\textbf{a)}, linewidths distribution of the 108(D) model of \mbox{$v$-SiO$_2$} computed explicitly on a $5{\times}5{\times}5$ $\bm{q}$-mesh (light blue is at 30 K, light green is at 200 K, and orange is at 1300 K) or computed at the point $\bm{q}{=}\bm{0}$ only (dark blue is at 30 K, dark green is at 200 K, and red is at 1300 K). 
  The purple region represents the overdamped regime, where vibrations cannot be accurately described using the Wigner formulation and spectral-function approaches have to be employed \cite{simoncelli2021Wigner,Caldarelli_2022}.
  The horizontal dotted line is the broadening $\hbar\eta$ used in the Voigt renormalization for the 108-atom model, all the linewidths below this line are regularized and thus have negligible effect on the rWTE conductivity. The dashed lines are the analytical functions $\Gamma_a[\omega]$, determined from the distributions at $\bm{q}{=}\bm{0}$ only as detailed in Eqs.~(\ref{eq:approx_analytical_f},\ref{eq:approx_analytical_f2}).
  Panel \textbf{b)} shows the bare WTE conductivity of the 108(D) model computed exactly (\textit{i.e.} using the linewidths explicitly computed on the $5{\times}5{\times}5$ $\bm{q}$-mesh, solid red), or using the linewidths approximatively determined using the function $\Gamma_a[\omega]$ (dashed blue);
  the solid green and dotted black lines show the rWTE conductivities computed using the exact or approximated linewidths, respectively.
  The good agreement between the exact and the approximated calculations shows that the approximation developed allows to account for anharmonicity at a reduced computational cost and without appreciably compromising accuracy, both in the WTE and rWTE calculations.
   }
  \label{fig:interpolation}
\end{figure}
Specifically, we show in Fig.~\ref{fig:interpolation}\textbf{a)} that the approximated functions $\Gamma_a[\omega]$ (dashed lines, different colors show different temperatures) captures the trend of the linewidths explicitly computed over a dense $5{\times}5{\times}5$ $\bm{q}$ mesh (dense distributions). We recall that $\Gamma_a[\omega]$ is determined from the linewidth distributions at $\bm{q}{=}\bm{0}$ only (coarse clouds of scatter points).
In Fig.~\ref{fig:interpolation}\textbf{b)} we show that computing the bare WTE conductivity over a dense $5{\times}5{\times}5$ $\bm{q}$ mesh using the anharmonic linewidths computed exactly (solid red) or determined approximatively using the function $\Gamma_a[\omega]$ (dashed blue) are practically indistinguishable; consequently, also the rWTE conductivity computed over the same $5{\times}5{\times}5$ $\bm{q}$ mesh is practically unchanged when the exact (solid green) or approximated (dotted black) linewidths are used.

\newpage
\section{Quantum harmonic  specific heat} 
\label{sec:quantum_harmonic_specific_heat}
\begin{figure}[h]
  \centering
  \includegraphics[width=\WidthFigure]{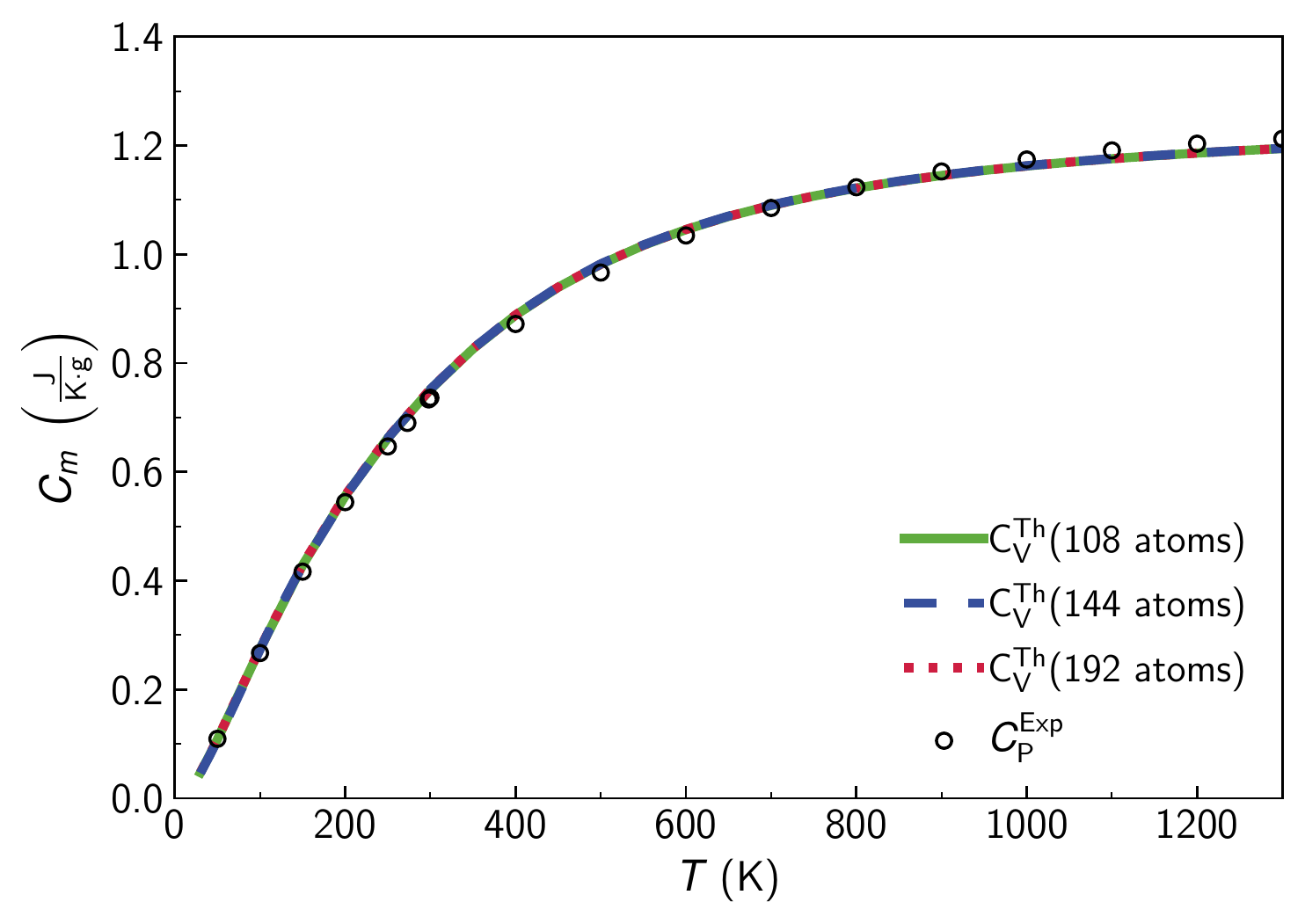}\\[-5mm]
  \caption{\textbf{Specific heat of vitreous silica.}  Lines are the quantum harmonic specific heat at constant volume and as a function of temperature computed from first principles for the 108(D) (green), 144(D)  (blue) and 192(D) (red) vitreous-silica models. Circles are measurements \cite{richet1982thermodynamic} of the specific heat at constant pressure.}
  \label{fig:spec_heat}
\end{figure}

We show in Fig.~\ref{fig:spec_heat} that the theoretical  quantum
 harmonic specific heat at constant volume  ($C_{\rm V}^{\rm Th}(T)=\frac{1}{\rho \mathcal{V}N_{\rm c}}\sum_{\bm{q},s} C(\bm{q})_{s}$, where $\rho$ is the density) is in close agreement with the experimental specific heat at constant pressure \cite{richet1982thermodynamic}. This suggests that the renormalization of vibrational energies due to anharmonicity and temperature \cite{PhysRevB.96.014111,aseginolaza2018phonon,PhysRevLett.125.085901,PhysRevB.102.201201}
 are negligibly small for $v$-SiO$_2$ in the temperature range considered \cite{horbach1999specific}, and as such these effects are not considered in this work.

\section{Velocity operator as a function of frequency} 
\label{sec:velocity_operator_as_a_function_of_frequency}
In this section we provide all the details on the plot reported in Fig.~\ref{fig:v_operator}.
In order to gain insights into the increasing, non-saturating trend of the conductivity with temperature (Fig.~\ref{fig:tc}), we recast the conductivity formula~(\ref{eq:thermal_conductivity_combined}) as a function of $\omega_{\rm a}{=}(\omega(\bm{q})_s{+}\omega(\bm{q})_{s'})/2$ and $\omega_{\rm d}{=}\omega(\bm{q})_s{-}\omega(\bm{q})_{s'}$:
\begin{equation}
\begin{split}
\kappa=&\int\limits_0^{\omega_{\rm max}}\hspace*{-1mm}d\omega_{\rm a}\hspace*{-1mm}\int\limits_{-\omega_{\rm max}}^{\omega_{\rm max}}\hspace*{-1mm}d\omega_{\rm d}
\Bigg[
\frac{1}{\mathcal{V}N_{\rm c}}{\sum_{\bm{q},s,s'}}
\frac{\omega(\bm{q})_s{+}\omega(\bm{q})_{s'}}{4}\frac{\rVert\tens{v}(\bm{q})_{s,s'}\lVert^2}{3}\\
\times&\left(\frac{C(\bm{q})_s}{\omega(\bm{q})_s}{+}\frac{C(\bm{q})_{s'\!}}{\omega(\bm{q})_{s'\!}}\right)\! 
\pi\mathcal{F}_{[\Gamma(\bm{q})_s{+}\Gamma(\bm{q})_{s'},\eta]}(\omega(\bm{q})_s{-}\omega(\bm{q})_{s'})\\
\times&{\delta}_{\sigma_{\rm a}}\hspace{-1mm}\left(\frac{\omega(\bm{q})_s{+}\omega(\bm{q})_{s'}}{2}{-}\omega_{\rm a}\right) 
{\delta}_{\sigma_{\rm d}}\hspace{-0.5mm}\big((\omega(\bm{q})_s{-}\omega(\bm{q})_{s'}){-}\omega_{\rm d}\big)
\Bigg]
\label{eq:thermal_conductivity_rewritten}
\raisetag{40mm}
\end{split}
\end{equation}
where the distributions $\delta_{\sigma}$ are Gaussian broadenings of the Dirac delta: 
\begin{equation}
  \delta_{\sigma}\left(\Omega-\omega\right)=\frac{1}{\sqrt{2\pi}\sigma} \exp\left[{-\frac{1}{2\sigma^2}\left(\Omega-\omega\right)^2}\right].
\end{equation}
In order to achieve our goal to recast Eq.~(\ref{eq:thermal_conductivity_rewritten}) in terms of physically-insightful frequency-dependent functions, we approximate the linewidths as single-valued function of frequency, \textit{i.e.} $\Gamma(\bm{q})_s=\mathcal{L}(\omega(\bm{q})_s)$; using this approximation allows to recast Eq.~(\ref{eq:thermal_conductivity_rewritten}) as follows:
\begin{equation}
\begin{split}
\kappa=\int\limits_0^{\omega_{\rm max}}d\omega_{\rm a}\int\limits_{-\omega_{\rm max}}^{\omega_{\rm max}}&d\omega_{\rm d}
\mathcal{G}(\omega_{\rm a},\omega_{\rm d})\mathcal{C}(\omega_{\rm a},\omega_{\rm d})
\left<|V^{\rm avg}_{\omega_{\rm a},\omega_{\rm d}}|^2\right>\\
\times\pi&\mathcal{F}_{[\mathcal{L}(\omega_{\rm a}+\frac{\omega_{\rm d}}{2}){+}\mathcal{L}(\omega_{\rm a}-\frac{\omega_{\rm d}}{2}),\eta]}(\omega_{\rm d}),
\label{eq:thermal_conductivity_rewritten2}
  \raisetag{5mm}
\end{split}
\end{equation}
where $\mathcal{G}(\omega_{\rm a},\omega_{\rm d})$ is a density of states
\begin{equation}
\begin{split}
  \mathcal{G}(\omega_{\rm a},\omega_{\rm d})
  =\frac{1}{N_{\rm at}}\frac{1}{\mathcal{V}N_{\rm c}}&{\sum_{\bm{q},s,s'}}{\delta}_{\sigma_{\rm a}}\left(\frac{\omega(\bm{q})_s+\omega(\bm{q})_{s'}}{2}-\omega_{\rm a}\right) \\
 \times&{\delta}_{\sigma_{\rm d}}\big((\omega(\bm{q})_{s}{-}\omega(\bm{q})_{s'})-\omega_{\rm d}\big)\;,
  \label{eq:2freq_vDOS}
  \raisetag{5mm}
\end{split}
\end{equation}
$\mathcal{C}(\omega_{\rm a},\omega_{\rm d})$ is a specific heat 
\begin{equation}
  \mathcal{C}(\omega_{\rm a},\omega_{\rm d})=\frac{\omega_{\rm a}}{2}\left(\frac{C(\omega_{\rm a}+\frac{\omega_{\rm d}}{2})}{\omega_{\rm a}+\frac{\omega_{\rm d}}{2}}+\frac{C(\omega_{\rm a}-\frac{\omega_{\rm d}}{2})}{\omega_{\rm a}-\frac{\omega_{\rm d}}{2}}\right),\;
  \label{eq:2freq_spec_heat}
\end{equation}
and $\left<|V^{\rm avg}_{\omega_{\rm a},\omega_{\rm d}}|^2\right>$ is the average square modulus of the velocity operator defined in Eq.~(\ref{eq:v_operator_omega_a_omega_d}) (whose Dirac delta must be broadened with the Gaussian $\delta_\sigma$ discussed here) and plotted in Fig.~\ref{fig:v_operator}.
Eq.~(\ref{eq:thermal_conductivity_rewritten2}), together with Fig.~\ref{fig:v_operator} and Fig.~\ref{fig:lw_compare}, sheds light on the saturating trend of the conductivity with temperature (Fig.~\ref{fig:tc}\textbf{a)}).
In fact, among the various quantities entering in Eq.~(\ref{eq:thermal_conductivity_rewritten2}),  the density of states~(\ref{eq:2freq_vDOS}) and the specific heat~(\ref{eq:2freq_spec_heat}) have a trivial temperature dependence (the former is temperature-independent, the latter saturates with increasing temperature). 
The change of variable performed in Eq.~(\ref{eq:thermal_conductivity_rewritten}) shows that the temperature-conductivity trend is determined by how the average square velocity-operator elements vary with the vibrational frequency difference $\omega_{\rm d}$, because the increase of the linewidths with temperature (Fig.~\ref{fig:lw_compare}\textbf{a)}) results in a broader Lorentzian distribution~(\ref{eq:Lorentzian}) that encloses velocity-operator elements corresponding to increasingly larger frequency differences $\omega_{\rm d}$.
In particular, for vitreous silica, the average square velocity-operator elements
 are almost independent from $\omega_{\rm d}$ for all values of $\omega_{\rm a}$ (Fig.~\ref{fig:v_operator}).
 It follows that the saturating trend of the conductivity at high temperature reported in Fig.~\ref{fig:tc}\textbf{a)} is inherited from the saturating trend of the specific heat. 

Fig.~\ref{fig:v_operator} is computed relying on the 192(D) model and on first-principles calculations, using a $3{\times}3{\times}3$ $\bm{q}$-point mesh, $\hbar\sigma_{\rm a}=15$ cm$^{-1}$ and $\hbar\sigma_{\rm d}=\sqrt{\frac{\pi}{2}}\hbar\eta=5$ cm$^{-1}$ (this latter corresponds to a Gaussian having height $\frac{1}{\pi\eta}$, with $\hbar\eta=4$ cm$^{-1}$ equal to the one used in the computation of the AF conductivity for the 192(D) model). Increasing the $\bm{q}$-point mesh to $5{\times}5{\times}5$ or multiplying $\sigma_{\rm a}$ and $\sigma_{\rm d}$ by a factor of 2 does not produce significant changes. We have verified that plotting the velocity operator for the other models studied from first principles or using GAP yields results that are practically indistinguishable from those reported in Fig.~\ref{fig:v_operator}. 

\section{Effects of anharmonicity using the BKS potential} 
\label{sec:effect_of_anharmonicity_using_the_bks_potential}
\begin{figure}[h]
  \centering
  \includegraphics[width=\WidthFigure]{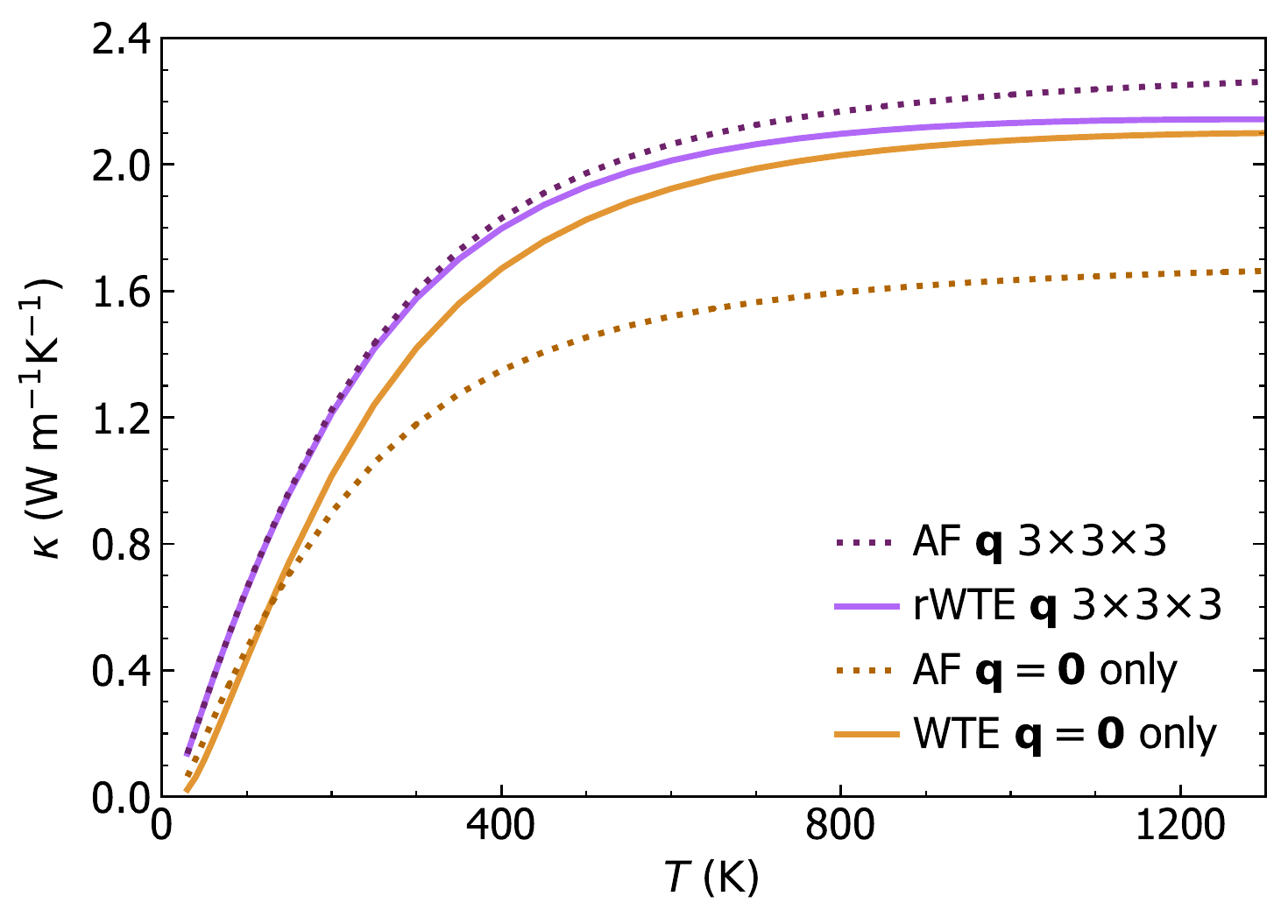}
  \caption{
  {\textbf{Negligible effects of anharmonicity in $v$-SiO$_2$ described with the BKS potential.}
 Evaluating the conductivity using the protocol to determine the bulk limit yields an anharmonicity rWTE conductivity (solid purple) that differs only about 4$\%$ from the bulk limit of the AF conductivity (dotted purple, evaluated using a Gaussian broadening $\hbar\eta=4\;{\rm cm}^{-1}$ determined from a convergence test analogous to that reported in Fig.~\ref{fig:harm_theory_plateau}).
 In contrast, evaluating the AF and WTE conductivities at $\bm{q}{=}\bm{0}$ only (and using a Lorentzian broadening $\hbar\eta=1.6\;{\rm cm}^{-1}$ in the former case) yield values that strongly underestimates the bulk value for the AF conductivity (dotted orange) and weakly underestimates the Wigner conductivity (solid orange), respectively.  }}
  \label{fig:BKS_anh}
\end{figure}

{While our BKS-based rWTE predictions are in agreement with Ref.\cite{PhysRevB.106.014305}, our BKS-based AF conductivity is significantly different from that reported in Ref. \cite{PhysRevB.106.014305}, which claimed that anharmonicity enhances appreciably  the conductivity in \mbox{$v$-SiO$_2$} (specifically, Ref.\cite{PhysRevB.106.014305} concluded that in \mbox{$v$-SiO$_2$}  in the high-temperature limit, the anharmonic WTE conductivity  is about $30\%$ higher than the AF conductivity). 
In Fig.~\ref{fig:BKS_anh} we show that evaluating the AF conductivity as done in Ref.\cite{PhysRevB.106.014305}, \textit{i.e.} using the BKS potential, at $\bm{q}=\bm{0}$ only and using a Lorentzian broadening $\hbar\eta=1.6\;{\rm cm}^{-1}$, yields a value that significantly underestimates the bulk value for the AF conductivity. 
Using the Fourier interpolation with an aptly chosen broadening to determine the bulk limit ($\hbar\eta=4\;{\rm cm}^{-1}$) yields a larger AF conductivity, which differs only about 4$\%$ from the bulk limit of the rWTE conductivity.}

\section{Calculations performed using the GAP potential} 
\label{sec:GAP_calculations}
The GAP potential for vitreous silica was taken from Ref.~\cite{erhard2022machine}. The 5184(G) and 192(G) models discussed in Figs.~\ref{fig:silica_AF_GAP},\ref{fig:harm_theory_plateau},\ref{fig:linewidths_sampling} are taken from Ref.~\cite{erhard2022machine}, we chose the 192-atom structure number 4 in that reference as the one having the closest density to the large 5184-atom model. 
More details on these structures are reported in Ref.~\cite{erhard2022machine}.
The densities of the models analyzed are reported in Tab.~\ref{tab:density}.
We have checked that the AF conductivity of the other four 192-atom models of $v$-SiO$_2$ discussed in Ref.~\cite{erhard2022machine} are very similar to the bulk AF conductivities shown in Fig.~\ref{fig:silica_AF_GAP} (differences are smaller than the $9\%$ differences discusses in Fig.~\ref{fig:other_models}).

Interatomic forces and stress tensor are computed using \textsc{lammps}\cite{brown2011implementing}, using the \textsc{quip}\cite{Csanyi2007-py,GAP_potential,Kermode2020-wu} interface to call the GAP potential routines. 
Cell parameters and atomic positions are relaxed using a threshold of $25$ eV/Angstrom for the atomic forces (\textit{i.e.} a structure is considered relaxed if all the Cartesian components of the forces acting on atoms are smaller than this threshold), and a threshold of 0.1 kBar for pressure.
The harmonic dynamical matrices (which yield the vibrational frequencies and velocity operators) are computed using \textsc{phonopy}, on a $2{\times}2{\times}2$ supercell for the 192(G) structure, and on the reference cell ($1{\times}1{\times}1$) for the  5184(G) model.
For the 192(G) model, third-order force constants are computed in the reference cell using ShengBTE \cite{li2014shengbte}, using  a cutoff equal to the $6^{\rm th}$ nearest-neighbor; the resulting force constants are then converted in \textsc{phono3py}\cite{phono3py} format using \textsc{hiphive}\cite{hiphive}.
The anharmonic linewidths for the 192(G)  model (shown in Fig.~\ref{fig:linewidths_sampling}) are computed using \textsc{phono3py} at $\bm{q}=\bm{0}$ and using Gaussian smearing of $\hbar\sigma=2\; {\rm cm}^{-1}$ for the Dirac delta appearing in the collision operator. 
The anharmonic linewidths for the 5184(G) model are determined using the approximation detailed in Fig.~\ref{fig:linewidths_sampling} and in Fig.~\ref{fig:interpolation} in the Methods.

\section{First-principles calculations} 
\label{sec:first_principles_calculations}
All the density-functional theory (DFT) calculations have been performed with the Quantum \textsc{espresso} distribution \cite{giannozzi2017advanced} using the PBE functional with Grimme-D2 corrections (PBE+D2) \cite{grimme2006semiempirical}. This choice is motivated by the benchmarks given in Ref.~ \cite{Mauri_choice_functional} and accounting for the capability of Quantum  \textsc{espresso} to compute phonons using density-functional perturbation theory (DFPT) \cite{RevModPhys.73.515} with the PBE+D2 exchange-correlation (XC) functional.
This choice is validated by the agreement between theoretical and experimental densities reported in table~\ref{tab:density} and also by the capability of this functional to accurately describe the thermal properties of $\alpha$-quartz (Fig.~\ref{fig:conductivity_quartz_cristobalite}). 
We have used pseudopotentials from the SSSP efficiency library \cite{prandini2018precision,lejaeghere2016reproducibility} with a cutoff of $50$ Ry and a dual of 8. In the following we report the details for all the three different systems studied: vitreous silica, $\alpha$-cristobalite, and $\alpha$-quartz.

\begin{table}[b]
  \caption{\textbf{Densities of silica polymorphs.} Comparison between theoretical and experimental  densities $\rho$ for the various silica polymorphs  analyzed.  }
  \label{tab:density}
  \centering

  \begin{tabular}{l|c}
  \hline

  \hline
  \textbf{Structure} & {$\rho$ ($\rm kg/m^{3}$)} \\
  \hline
    $v$-SiO$_2$, 192(D) PBE+D2 (SiO2 2818 \cite{Kroll_2013})  & 2241.2\\
    $v$-SiO$_2$, 192(D) GAP (SiO2 2818 \cite{Kroll_2013})  & 2288.6\\
    $v$-SiO$_2$, 192(D) BKS (SiO2 2818 \cite{Kroll_2013})  & 2243.8\\
\hline
    $v$-SiO$_2$, 192(G) GAP ($\#4$ \cite{erhard2022machine})  & 2188.7\\
    $v$-SiO$_2$, 5184(G) GAP \cite{erhard2022machine}  & 2203.6 \\
  \hline  
    $v$-SiO$_2$, 144(D) PBE+D2 \cite{PhysRevB.79.064202}             & 2220.6 \\
    $v$-SiO$_2$, 108(D) PBE+D2 (SiO2.1586 \cite{charpentier2009first}) & 2243.9\\
  \hline
$v$-SiO$_2$, Experiment \cite{brueckner1970properties}     & 2203 $\pm$ 3\\
$v$-SiO$_2$, Experiment \cite{Heraeus2010}     & 2200  $\pm$ 10  \\
  \hline
  \hline
$\alpha$-cristobalite \cite{cristobalite_structure_cif}, PBE+D2 & 2383.5 \\
\hline
$\alpha$-cristobalite, Experiment \cite{downs1994pressure} & 2326$\pm$12 \\

  \hline
  \hline
$\alpha$-quartz \cite{quartz_structure}, PBE+D2 & 2641.9 \\
\hline
$\alpha$-quartz, Experiment \cite{2014crc} & 2650 \\
  \hline 
  \end{tabular}
\end{table}

\textit{Vitreous silica.}
The 108(D) structure of vitreous silica is taken from Ref. \cite{charpentier2009first} (``SiO2.1586'' model).
For this structure, cell parameters and atomic positions are relaxed with DFT using $\Gamma$-point ($\bm{q}{=}\bm{0}$) sampling, a threshold of $10^{-4}$ Ry/Bohr for the atomic forces (\textit{i.e.} a structure is considered relaxed if all the Cartesian components of the forces acting on atoms are smaller than this threshold), and a threshold of 0.1 kBar for pressure.
The harmonic dynamical matrices (which yield the vibrational frequencies and velocity operators) are computed using DFPT on a $2{\times}2{\times}2$  $\bm{q}$-point mesh  and accounting for the non-analytic term correction due to the dielectric tensor and Born effective charges.
Third-order force constants are computed in the reference cell using ShengBTE \cite{li2014shengbte}, and with a cutoff of $0.32$ nm (corresponding to the $6^{\rm th}$ nearest-neighbor). 
The linewidths are computed using \textsc{phono3py} \cite{phono3py} on a $5{\times}5{\times}5$ $\bm{q}$-point mesh 
using the tetrahedron method, we checked that computing them using Gaussian smearing of $\hbar\sigma=2\; {\rm cm}^{-1}$ for the Dirac delta appearing in the collision operator does not produce appreciable changes. Thermal conductivity calculations are performed using a $\bm{q}$-interpolation mesh equal to $5{\times}5{\times}5$. 
The Voigt profile, used to combine the AF and the WTE conductivities as detailed before, has been numerically implemented following the prescriptions reported in Ref. \cite{ida2000extended}. 
{We have checked that reducing the $\bm{q}$-interpolation mesh  to $3{\times}3{\times}3$ does not produce appreciable changes.}

The 192(D) model is generated using the same techniques discussed in Ref. \cite{charpentier2009first} and is discussed in Ref.~\cite{Kroll_2013}, while the 144(D) model is taken from Ref. \cite{PhysRevB.79.064202} and is available in the Materials Cloud Archive \cite{Materials_cloud_structure}.
For all these structures, the relaxation of the reference cell with DFT, the calculation of the harmonic dynamical matrices, and the the calculation of third-order anharmonic force constants are performed using the same parameters used for the 108-atom structure. 
These harmonic and anharmonic force constants are then used to compute the linewidths at the point $\bm{q}=\bm{0}$ only using \textsc{phono3py} \cite{phono3py} and a Gaussian smearing of $2\;{\rm cm^{-1}}$, and then employed within the approximated procedure discussed and validated in Fig.~\ref{fig:interpolation}. 
We also verified that the 192(D)  model accurately reproduces the experimental bulk modulus of $v$-SiO$_2$ (the theoretical value we computed is 36.9 GPa, while the experimental value of Ref. \cite{deschamps2014elastic} is 36.8 GPa and the experimental value of Ref. \cite{kondo1981nonlinear} is  36.9 GPa). 

An analysis of the coordination numbers using the procedure based on the minimum of the radial distribution function as implemented in the \textsc{r.i.n.g.s.} software \cite{le2010ring} or that based on the position of the Wannier centers discussed in Ref. \cite{silvestrelli1998maximally} have revealed in both cases that the 108(D), 144(D), 192(D) vitreous structures considered in this work do not have coordination defects or lone pairs (both before and after the DFT relaxation).\\

\textit{$\alpha$-cristobalite.}
The crystal structure of  $\alpha$-cristobalite is taken from Ref. \cite{cristobalite_structure_cif} (ICSD collection code 47219).
In first-principles calculations, the Brillouin zone is integrated with a Monkhorst–Pack mesh of $5{\times}5{\times}4$ points, with a (1, 1, 1) shift.
Second-order force constants are computed using DFPT on a $5{\times}5{\times}4$ $\bm{q}$-point mesh,
 accounting also for the non-analytic term correction due to the dielectric tensor and Born effective charges.
To obtain the third-order force constants the finite-difference method implemented in ShengBTE \cite{li2014shengbte} is used, together with the interconversion software from ShengBTE to Quantum \textsc{espresso}, available in the \textsc{d3q} package \cite{fugallo2013ab,paulatto2015first}. In these third-order force constants calculations, a $2{\times}2{\times}2$ supercell with a $2{\times}2{\times}2$ k-point sampling is used, and interactions up to the $6^{\rm th}$ nearest neighbor (corresponding to $\sim 0.39$ nm) are considered.
The linewidths (Fig.~\ref{fig:lw_compare}\textbf{b)}) and thermal conductivity (Fig.~\ref{fig:conductivity_quartz_cristobalite}\textbf{a)}) are computed with the \textsc{d3q} package \cite{fugallo2013ab,paulatto2015first} using a $17{\times}17{\times}13$ $\bm{q}$-point mesh and a Gaussian smearing $\hbar\sigma=4\;{\rm cm}^{-1}$.\\

\textit{$\alpha$-quartz.}
The crystal structure of $\alpha$-quartz is taken from Ref. \cite{quartz_structure} (Crystallographic Open Database id 1526860).
In first-principles calculations, the Brillouin zone is integrated with a Monkhorst–Pack mesh of $6{\times}6{\times}5$ points, with a (1, 1, 1) shift.
Second-order force constants are computed using DFPT on a $4{\times}4{\times}4$ $\bm{q}$-point mesh,  accounting also for the non-analytic term correction due to the dielectric tensor and Born effective charges.
Third-order force constants are computed using the finite difference methods as implemented in ShengBTE \cite{li2014shengbte}, using a $3{\times}3{\times}2$ supercell with a $\Gamma$-only k-point sampling, and a cutoff for atomic interactions of $0.9$~nm.
The linewidths (Fig.~\ref{fig:lw_compare}\textbf{c)}) and thermal conductivity (Fig.~\ref{fig:conductivity_quartz_cristobalite}\textbf{b)}) are computed with the \textsc{d3q} package \cite{fugallo2013ab,paulatto2015first}  using a $19{\times}19{\times}15$ $\bm{q}$-point mesh and a Gaussian smearing $\hbar\sigma=4\;{\rm cm}^{-1}$. 
Results are compatible to those reported in Ref. \cite{Togo_quartz_2018}.

We conclude by noting that in Fig.~\ref{fig:DOS_diff_spec_heat} the vDOS for all materials are computed using the same meshes used for the thermal conductivity calculations detailed above, and a Gaussian smearing of 15 cm$^{-1}$ for the vibrational energies.\\

\noindent
{{\textbf{{Data availability.}}}}  Raw data were generated using the SCITAS High Performance Computing facility at the {\'E}cole Polytechnique F{\'e}d{\'e}rale de Lausanne and the Cambridge Service for Data-Driven Discovery (CSD3). 
The atomistic models of vitreous silica, $\alpha$-quartz, and $\alpha$-cristobalite studied in this work are available on the Materials Cloud Archive \cite{Materials_cloud_release,materialsCloud_ref}.\\[3mm]

\noindent
{{\textbf{{Code availability.}}}} Quantum \textsc{espresso}\cite{giannozzi2017advanced} is available at \url{www.quantum-espresso.org}; the scripts related to the computation of the third order force constants using the finite-difference method are available at \url{bitbucket.org/sousaw/thirdorder}; 
\textsc{phonopy} and \textsc{phono3py} are available at \url{github.com/phonopy}.
The GAP potential for silica polymorphs is available at \cite{erhard2022machine}. \textsc{lammps}\cite{brown2011implementing} is available at \url{www.lammps.org} and the interface for \textsc{lammps} with the GAP potential\cite{Csanyi2007-py,GAP_potential,Kermode2020-wu}  is available at \url{github.com/libAtoms/QUIP}.


\end{document}